\documentclass[numberedappendix,iop]{emulateapj}
\usepackage{amsmath}
\usepackage{mathtools}
\usepackage{graphicx}
\usepackage{changes}
\usepackage{color}
\usepackage{soul} 

\newcommand{\ie}{\emph{i.e.} }
\newcommand{\eg}{\emph{e.g.,} }

\newcommand{\be}{\begin{equation}}
\newcommand{\ee}{\end{equation}}
\newcommand{\bea}{\begin{equation*}}
\newcommand{\eea}{\end{equation*}}
\newcommand{\beqr}{\begin{eqnarray} \nonumber}
\newcommand{\eeqr}{\end{eqnarray}}
\newcommand{\beqrb}{\begin{eqnarray}}
\newcommand{\eeqrb}{\nonumber \end{eqnarray}}
\newcommand{\fin}{\mbox{ .}}
\newcommand{\coma}{\mbox{ ,}}

\newcommand{\km}{\mbox{ km}}

\newcommand{\pasa}{Publications of the Astron. Soc. of Australia}

\shorttitle{Phase Space Analysis of Core-Collapse Supernova}

\newcommand{\fig}[1]{{#1}}

\newcommand{\fixme}[1]{{}}

\newcommand{\RS}{{R_S}}
\newcommand{\RP}{{R_P}}
\newcommand{\VS}{{V_S}}

\begin{document}
\title{Shock Revival in Core-Collapse Supernovae: A Phase-Diagram Analysis} 
\author{Daniel Gabay and Shmuel Balberg}
\affil{Racah Institute of Physics, The Hebrew University of Jerusalem, Edmond J. Safra Campus, Jerusalem 91904, Israel}
\and
\author{Uri Keshet}
\affil{Physics Department, Ben-Gurion University of the Negev, PO Box 653, Be'er-Shvea 84105, Israel}

\begin{abstract}
We examine the conditions for the revival of the stalled accretion shock in core-collapse supernovae, in the context of the neutrino heating mechanism. We combine one dimensional simulations of the shock revival process with a derivation of a quasi-stationary approximation, which is both accurate and efficient in predicting the flow. In particular, this approach is used to explore how the evolution of the system depends on the shock radius, $R_S$, and velocity, $V_S$ (in addition to other global properties of the system). We do so through a phase space analysis of the shock acceleration, $a_S$, in the $R_S-V_S$ plane, shown to provide quantitative insights into the initiation of runaway expansion and its nature. In the particular case of an initially stationary ($V_S=0,\;a_S=0$) profile, the prospects for an explosion can be reasonably assessed by the initial signs of the partial derivatives of the shock acceleration, in analogy to a linear damped/anti-damped oscillator. If $\partial a_S/\partial R_S<0$ and $\partial a_S/\partial V_S>0$, runaway expansion will likely occur after several oscillations, while if $\partial a_S/\partial R_S>0$, runaway expansion will commence in a non-oscillatory fashion. These two modes of runaway correspond to low and high mass accretion rates, respectively. We also use the quasi-stationary approximation to assess the advection-to-heating timescale ratio in the gain region, often used as an explosion proxy. Indeed, this ratio does tend to $\sim1$ in conjunction with runaway conditions, but neither this unit value nor the specific choice of the gain region as a point of reference appear to be distinct conditions in this regard.
\end{abstract}

\section{Introduction}

The physical mechanism which drives core collapse supernovae remains an outstanding problem after several decades of research. While there exists clear evidence that massive stars do explode \citep{Smartt2009}, a viable explosion mechanism has not yet been demonstrated robustly by theoretical means (for recent reviews, see \citet{Burrows2013,Janka2012,Foglizzoetal2015}).

Since first proposed by \citet{wilson1985} and \citet{bethe_wilson1985}, "delayed neutrino heating" has generally been considered the most plausible mechanism for driving core collapse supernovae (see however, \citet{Kushnir2015} for an alternative view). The process envisioned for this mechanism is essentially two-staged. First, the iron core of the massive star collapses to a proto-neutron-star (PNS) which is stabilized by the strong nuclear force, creating an accretion shock that plows the incoming material which flows inward. This shock stalls, due to heavy energy losses in neutrino cooling of the shocked accretion layer (and also through dissociation of nuclei as they cross the shock front), but is eventually revived when heating of this layer by neutrinos emitted from the PNS overcomes the energy losses and the inertia of the incoming material. The fundamental issue regarding this mechanism is how (and if!) the competition between neutrino heating and cooling, as well as gravity and ram pressure, can  revive the accretion shock and drive it into an outgoing expansion, eventually disrupting the entire envelope of the star.

After three decades of simulations of neutrino heating following core collapse, the overall picture is still a complicated one. There is a broad consensus that self-consistent, one dimensional simulations generally fail to explode for all but the lowest-mass progenitors \citep{Liebendorferal2001,Kitauraetal2006}. This fact has motivated a shift towards two- and subsequently three-dimensional (3D) simulations. In these simulations, multi-dimensional effects, such as turbulence, convection, rotation and instabilities, come into play, and have been demonstrated to generally aid an explosion (see, e.g., \citet{Couch2013,Fernandez2015,Melsonetal2015}). On the other hand, state of the art 3D simulations have yet to resolve the long standing problem: they tend to predict sub-energetic explosions (kinetic energies of a few $10^{50}\;$erg or less), or no explosion at all (see the discussion in \citet{Burrows2013} and references therein). However, the complexity of the neutrino-driven mechanism (diverse in both the physical processes and the wide range of distances and time scales involved) means that further advances may yet change this conclusion (see, e.g., \cite{Lentzetal2015} for a more favorable outlook).

This inherent intricacy has also led to an additional class of studies: effective simulations and calculations, which use simplified assumptions and physics. Such works compromise on quantitative accuracy in order to facilitate a qualitative understanding and a more straightforward parameter survey for identifying the underlying principles which are necessary to generate an explosion. A cornerstone of this line of research has been to invoke the neutrino luminosity as a free parameter in the simulations (a "neutrino light-bulb"), instead of generating it self-consistently in the simulations. Unsurprisingly, a fiducial increase of the neutrino luminosity does lead to an explosion. These analyses gave rise to the concept of a "critical neutrino luminosity", which is the minimal luminosity that drives a runaway explosion when all other parameters of the system are predetermined. The critical luminosity was first introduced by \citet{burrows_goshy93}, who considered the problem in the stationary approximation, identifying as critical the minimum luminosity for which no steady-state solution exists.

Stationary models of similar nature were applied also in recent works, in search of the origin of the critical nature of the problem, as well as a clearer condition for an explosion (see \citet{pejcha_thompson12,keshet_balberg12, Murphy_Dolence2015} and references therein).
For example, under simplifying assumptions such as a neutrinosphere of black body temperature $T_\nu=4T_4\mbox{ MeV}$ and mass density $\rho_\nu=10^{11}\rho_{11}\mbox{ g cm}^{-3}$, the critical luminosity for which a solution cannot be found is \citep{keshet_balberg12},
\begin{equation}
L_c \simeq 6.9 \times 10^{52} \Bigg[\frac{M_{P,1.3} T_4}{1+0.12\ln\left(\frac{M_{P,1.3}^2\rho_{11}}{\dot{M}_1 T_4^{3/2}}\right)} \Bigg]^2 \mbox{ erg s}^{-1} \, ,
\end{equation}
where $\dot{M}_1=\dot{M}/(1M_\odot \mbox{ s}^{-1})$ and $M_{P,1.3}\equiv M_P/1.3M_\odot$ are the normalized mass accretion rate and mass of the PNS.

Treating the neutrino luminosity as a free parameter in simulations has also led to the key observation that the critical luminosity in multi-dimensional simulations is lower than in one dimensional simulations under similar conditions \citep{murphy_burrows08,couchocconor2014}, thus explicitly highlighting the importance of multi-dimensional effects. We note that there are conflicting results regarding whether the critical luminosity in three dimensions is higher than in two dimensions \citep{bruennetal2009,nordhaus10,dolenceetal2013,Takiwakietal2014,couchocconor2014}.

In this work we revisit the critical nature of the transition from a steady accretion to a runaway explosion, with the aid of effective, one dimensional simulations. While some important multi-dimensional features are necessarily discarded when using this approach, it still qualitatively describes much of the dynamics which dictates the evolution of the accretion shock and the shocked material. By nature, one-dimensional simulations are better suited for parameter surveys, being easier to evaluate as a quantitatively well-defined problem. We are specifically motivated by the fact that in simulations, explosions can occur after the shock goes through a series of increasingly strong oscillations, rather than directly accelerating from a standing shock to runaway expansion \citep{onishietal06,murphy_burrows08,Fernandez2012}; naturally, this feature cannot be assessed by purely stationary models (in which the stability of the solution can only crudely be investigated \citep{nagakura13}). Our general goal is to examine which quantitative aspects of the flow determine the transition from an oscillating accretion shock to an explosion, and their relation to the critical neutrino luminosity. Furthermore, the critical neutrino luminosity for an explosion initiated by oscillations appears to be somewhat lower than that predicted by the stationary models, and we aim to uncover the reason for this trend.

The structure of this paper is as follows. In section \ref{sec:physical_model} we review the spherically-symmetric physical model used in this work. Our simulation code, developed originally for this work, is described in section \ref{sec:hydrodynamic_code}. Typical results concerning the accretion flow which transitions from oscillations to an explosion are shown in section \ref{sec:Flow}. 
Our fundamental observation, considering the conditions for a positive shock acceleration, is presented in section \ref{sec:shock_acceleration}.
Here, with the aid of appendices \ref{app:d2Idt2} and \ref{app:P_P},  we develop a quasi-stationary approximation which allows us to study the shock properties in the phase-space of the shock radius and velocity. We compare our conclusions to a frequently suggested timescale criterion for an explosion in section \ref{sec:explosion_condition}. In section {\ref{sec:conclusions} we summarize our conclusions and discuss some of their implications.
The oscillation period is estimated in appendix \ref{app:t_osc}.

\section{The Spherical Model} \label{sec:physical_model}
Simulations generally indicate that following core bounce, there is a transient phase of order 100 ms, after which the incoming mass accretion rate, $\dot{M}_0$, and the neutrino luminosity, $L$, settle on roughly fixed values during the evolution of the shock \citep{burrows_livne07,marek_janka09}. Correspondingly, it is common practice in simplified models of the explosion process to set these two parameters as constants, and study the dynamical dependence on their values. Here, we reduce the problem to an idealized spherically symmetric flow, as has been done in many similar studies (see, e.g., \citet{Fernandez2012}).
In spherical symmetry, the equations of motion, used to calculate the dynamics, are the conservation of mass, momentum and energy:
\begin{equation} \label{eq:con_density}
\frac{\mathrm \partial \rho}{\mathrm \partial t} + \frac{1}{r^2} \frac{\mathrm \partial }{\mathrm  \partial r} \left( r^2 \rho u \right) =0 \coma
\end{equation}
\begin{equation} \label{eq:con_momentum}
\frac{\mathrm \partial \left( \rho u \right) }{\mathrm \partial t} + \frac{1}{r^2} \frac{\mathrm \partial }{\mathrm  \partial r} \left( r^2 \rho u^2 \right) = - \frac{\partial p}{\partial r} - \frac{G M_P  \rho}{r^2} \coma
\end{equation}
and
\begin{equation} \label{eq:con_energy}
\frac{\mathrm \partial  \left( \rho e_{tot} \right) }{\mathrm \partial t} + 
\frac{1}{r^2} \frac{\mathrm \partial }{\mathrm  \partial r} \left[ r^2 \rho u \left(e_{tot}+\frac{p}{\rho} \right) \right] =\rho \dot{q} \coma
\end{equation}
where $u$, $\rho$, and $p$ are respectively the fluid radial velocity (in the lab, i.e. the PNS, rest frame), mass density, and pressure, and
\begin{equation}
e_{tot}=\frac{1}{2}u^2+e-\frac{G M_P}{r} 
\end{equation} 
is the specific total energy, and $e$ is the specific internal energy.
The net neutrino deposition rate (heating minus cooling) per unit fluid mass is denoted by $\dot{q}$. 

In the gravitational terms ($G$ being Newton's constant) $M_P$ is the PNS mass. In the general case, $M_P$ should be replaced with the mass enclosed within a radius $r$ at time $t$, but since in realistic scenarios the mass of the PNS dominates over the mass of the accretion layer, we neglect this layer's self-gravity and use a fixed central mass.

\subsection{The Basic Physical Model} \label{section:basic_model}

Equations (\ref{eq:con_density}-\ref{eq:con_energy}) are solved for an accretion layer located between the PNS and the shock radius, $\RS$. The immediate downstream of the shock is thus used as an outer boundary. It is common practice in simplified simulations of neutrino driven explosions to also specify an inner boundary at a radius identified with the PNS surface, $\RP$. Physically speaking, the radius $\RP$ roughly corresponds to the neutrinosphere, above which the neutrino luminosity is approximately constant. 

The outer boundary conditions at $\RS$ are determined by approximating that the preshocked material as in a pressure-less free-fall with a constant mass accretion rate $\dot{M}_0 = 4 \pi R^2_S \rho_0 u_0$. The velocity (in the lab frame, \ie the PNS frame) and density of infalling material at the shock are then
\begin{equation} \label{eq:free-fall}
u_0 = - \alpha \left(\frac {G M_P}{R_S}\right)^{1/2} , \quad\rho_0=\frac{|\dot{M}_0|}{4\pi \alpha (G M_P)^{1/2}} R_S^{-3/2} ,
\end{equation}
where $\alpha$ is a constant of order unity ($\sqrt{2}$ for perfect free-fall). The properties of the material in the immediate downstream of the shock are then determined by the Rankine-Hugoniot jump conditions,
\begin{equation} \label{eq:RH_density}
\rho_1 v_1 = \rho_0 v_0 \coma
\end{equation}
\begin{equation}\label{eq:RH_momentum}
P_1 + \rho_1 v_1^2 = p_0 + \rho_0 v_0^2 \coma
\end{equation}
and 
\begin{equation} \label{eq:RH_energy}
\frac{1}{2} v_1^2 +e_1+\frac{P_1}{\rho_1} = \frac{1}{2} v_0^2 +e_0+\frac{P_0}{\rho_0} - q_d \fin
\end{equation}
Here indices $0$ and $1$ denote quantities upstream and downstream of the shock, respectively, and $v=u-\VS$ is the fluid velocity relative to the shock, taking into account the (lab frame) shock velocity, $\VS$. Hereafter we denote $\rho_1\equiv \rho_S$, $u_1 \equiv u_S$  and $P_1\equiv P_S$, indicating the mass density, fluid (lab frame) velocity, and pressure just beneath the shock.

Finally, $q_d$ in equation (\ref{eq:RH_energy}) is the energy removed per unit mass through the dissociation of infalling ions by the shock.
The value of $q_d$ has several important consequences for the entire profile of the accretion layer. In terms of the boundary conditions, it determines the compression ratio across the shock,
$\beta=\rho_1/\rho_0$, given by
\begin{equation} \label{eq:beta_factor}
\beta =\frac{\gamma+1}{\gamma-\sqrt{1+2(\gamma^2-1)\theta}} \, \quad \mathrm{where} \quad  \theta\equiv\frac{q_d}{(u_0-\VS)^{2}}\;.
\end{equation}

\vspace{0.5cm}
Here, $\gamma=P/(\rho e)+1$ is the effective adiabatic index of the shocked material, which is typically radiation-dominated and so $\gamma \sim 1.4$ \citep{jan01}.
In the limit of zero dissociation $(q_d\rightarrow0, \theta\rightarrow0)$, the compression factor equals that of a simple strong shock, $\beta\rightarrow(\gamma+1)/(\gamma-1)$, but if dissociation is significant (compared to the kinetic energy of the infalling material in the shock reference frame) the compression factor will be larger.

\subsection{Further Simplifications} \label{section:effective_models}

As our goal in this work is a qualitative interpretation of the conditions for shock revival through neutrino heating, we apply some further simplifications to our model, which allow for a clearer insight into the numerical simulations. First, we use a simplified equation of state for the shocked material, describing radiation, nonrelativistic nucleons, and relativistic electrons of zero chemical potential and degeneracy \citep{yamasaki_yamada05,keshet_balberg12}:
\begin{equation}
p=\frac{11}{12} a T^4+\frac{k_B \rho T}{m_n}  \quad ,\quad   e=\frac{11}{4} \frac{a T^4}{\rho} +\frac{3}{2} \frac{k_B T}{m_n}\;,
\end{equation}
where $a$, $k_B$, and $m_n$ are the radiation constant, Boltzmann constant, and nucleon mass, respectively. While this equation of state neglects the finer aspects of the composition of the shocked material, it allows for more efficient simulations while still capturing the main essence of the problem, especially the transition from a radiation dominated state $(\gamma\approx4/3)$ near the shock to a matter dominated one $(\gamma\approx5/3)$ near the PNS, as the density increases by several orders of magnitude. Correspondingly, the dissociation energy, $q_d$, is treated as a free parameter, which can range between the non-dissociation limit $q_d=0$ and full dissociation of iron ions into free nucleons with $q_d=q_{Fe}\equiv8.5 \times 10^{18}\;\mathrm{erg \ g^{-1}}$. In reality, the actual dissociation across the shock is partial, due to recombination processes of nucleons into $\alpha$ particles, so that some of the dissociation energy is later added back to the accreted material.
Notice that when dissociation is included, there is a physical upper limit on the radius of the stagnation shock for free-falling material, $R_S(t=0) \lesssim 200M_{P,1.3}(q_d/q_{Fe})^{-1} \km$.

We also use a simple recipe for neutrino heating and cooling, which is often applied in simplified analytical and numerical models \citep{murphy_burrows08,jan01}. The simple formula for the total neutrino heating, $\dot{q}_H$ and cooling, $\dot{q}_C$, are
\begin{equation} \label{eq:heating}
\dot{q}_H = 1.54 \times 10^{20} L_{52} \left( \frac{r}{100\, \mathrm{km}} \right)^{-2} \left( \frac{T_{\nu_e}}{4\, \mathrm{MeV} } \right)^2 \mathrm{erg} \,\mathrm{g}^{-1}\,  \mathrm{s}^{-1}
\end{equation}
and
\begin{equation} \label{eq:cooling}
\dot{q}_C = 1.40 \times 10^{20} \left( \frac{T}{2\, \mathrm{MeV} } \right)^6 \mathrm{erg} \,\mathrm{g}^{-1}\, \mathrm{s}^{-1}
\end{equation}
where $L_{52}$ is the electron-neutrino luminosity in units of $10^{52} \mathrm{erg \; \ s^{-1}}$, $T_{\nu_e}$ is the electron-neutrino temperature at the neutrinospehre, and $T$ is the (radius-dependent) temperature of the matter and photons, assumed to be in equilibrium. The total energy deposition rate in equation (\ref{eq:con_energy}) is then $\dot{q}=\dot{q}_H - \dot{q}_C$. In equation (\ref{eq:heating}) it is assumed that the electron antineutrino luminosity is equal to the electron neutrino luminosity, and that the contribution of the other neutrino types to heating can be neglected. The total energy deposition rate in equation (\ref{eq:con_energy}) is then $\dot{q}=\dot{q}_H - \dot{q}_C$. 

Finally, since we are only considering matter above the neutrinosphere, where the optical depth is small, we assume a fixed neutrino luminosity above radius $\RP$, so that $L$ in Equation (\ref{eq:heating}) is independent of radius. In principle, the heating term can be corrected by a factor of $e^{-\tau}$, where $\tau(r)$ is the optical depth between radius $\RP$ and $r$, but we find that applying this correction has a minor effect on the bulk properties of the accretion layer, which are the focus of our present work. Therefore, we neglect this correction.

\section{The Simulation Code} \label{sec:hydrodynamic_code}

We solve the flow equations (\ref{eq:con_density}-\ref{eq:con_energy}) with the aforementioned boundary conditions and simplifications using a one-dimensional Lagrangian code. The code implements a standard von Neumann and Richtmyer staggered mesh method \citep{vonNeumann,richtmyer} for the equations of motion. The energy equation is solved implicitly with non-adiabatic contributions given by equations (\ref{eq:heating}-\ref{eq:cooling}).

The outer boundary is followed by continuously adding Lagrangian cells above the shock in free-fall velocity according to equation (\ref{eq:free-fall}). The shock dynamics are calculated with a quadratic artificial-viscosity, $\sigma$, added to the pressure term of the flow, but this requires some additional caution. Energy losses by dissociation at the shock (when included) can amount to a significant fraction of the internal energy of the post shock material, and as a result, numerical disturbances may arise within the artificial-viscosity scheme in which the shock is smoothed over a few grid cells. We circumvent this problem by applying the energy loss gradually as the cell passes through the shock region. For every cell entering the shock (with $\sigma \ge 0.5P$), the density and internal energy are saved, and the asymptotic post-shock compression factor, $\beta$, is calculated according to the density profile of the flow. We regulate the amount of energy lost by a cell, $\tilde{q}_d$, so that it reaches $q_d$ only after the cell is compressed to the asymptotic value. Quantitatively, $\tilde{q}_d$ is calculated by
\begin{equation} \label{eq:q_d_star}
\tilde{q}_d \left( \rho \right) = 
q_d \times \begin{cases}
1 & \mbox{ if } \rho>\rho_1 \, ; \\
\frac{\rho - \rho_0}{\rho_1 - \rho_0}  & \mbox{ if } \rho_0 < \rho\leq \rho_1 \, ; \\
0 & \mbox{ if } \rho\leq\rho_0 \, ,
\end{cases}
\end{equation}
where $\rho_0$ is the pre-shock density of the element, and $\rho_1=\beta \rho_0$ is the expected density after the shock. The cell then loses up to $\tilde{q}_d$ energy, but only as long as the internal energy does not drop below its value before entering the shock. This recipe guarantees that the internal energy in a cell cannot drop to negative values through rapid dissociation losses.

In order to ensure stability in the shock downstream, non-adiabatic processes (neutrino cooling and heating) are incorporated only after the cell has lost a total energy of $q_d$. This gradual dissociation loss recipe over a typical shock width (a few cells) is a minor correction since neutrino heating and cooling are generally much weaker near the shock than farther downstream, and we find that applying changes the critical luminosity by no more than a few percents, while guarantying numerical stability.

At the inner boundary, mass elements that enter the PNS (having $u<0$ at $R_P$) interact with a constant pressure $P_P$ in a ghost cell just below $R_P$, until they drop completely below $R_P$ (are absorbed in the PNS) or alternatively attain a positive velocity. The latter occurs only when the flow is well into a runaway expansion.

The simulations were typically calculated with about 500-1000 cells in the accretion layer, maintaining a decreasing cell width toward the PNS: the cell widths are adjusted so that each cell is thinner by a factor of $(1+\Delta)^{-1}$ with respect to its upper neighbor, with $\Delta \approx 10^{-2}-10^{-3}$. Rezoning is applied when necessary. We find that this resolution allows for numerical convergence, both in terms of the flow near the shock and in the steep density gradient near the PNS. 

\section{Features in an Explosive Flow} \label{sec:Flow}

We initialize a simulation for a given combination of $L, M_P, \RP$ and $\dot{M}_0$ by determining a stationary profile, i.e, solving the flow equations (\ref{eq:con_density}-\ref{eq:con_energy}) when all partial time derivatives are set to zero, as is the shock velocity, $\VS(t=0)=0$. The outer boundary conditions follow by setting a specific shock radius, $\RS(t=0)$, so we may solve the entire profile by integrating the stationary flow equations from the shock radius inward to $\RP$. The resulting pressure, $P_P$ in the ghost simulation cell just below $\RP$, is then determined, and subsequently serves as a time-independent inner boundary condition at $\RP$. As mentioned above, this pressure is assumed to be a property of the (unsimulated) PNS, and is thus kept fixed during the simulation. The dynamical evolution is then initiated with some small perturbation, either by intentionally shifting the shock radius (typically by few hundred meters), or simply by numerical noise. We find that as long as the initial perturbation is small, the evolution that follows does not depend on the specifics.

We mostly varied $L$ and $\dot{M}_0$ while keeping the PNS mass fixed at $M_P=1.3 M_\odot$, and its radius at $\RP=40 \; \mathrm{km}$. We also varied the dissociation parameter ($q_d$), between the non-dissociation limit $q_d=0$ and full dissociation of iron with $q_d=q_{Fe}$. The choice of the initial radius of the stalled shock $\RS(t=0)$ warrants some discussion. Without enforcing some additional constraint, this radius has no unique value. In the stationary approximation \citep{burrows_goshy93,pejcha_thompson12,keshet_balberg12}, the shock radius is uniquely determined, usually by requiring that the optical depth between the PNS and the shock be equal to $2/3$. As shown by \citet{keshet_balberg12}, for such an optical depth the shock radius is a slowly varying function of $L$ (when other parameters are kept fixed), and is typically  ($100-250) \; \mathrm{km}$. In the dynamic simulations we follow here, the accretion layer goes through a range of shock radii and optical depths, and so we opt to begin all simulations in a specific parameter survey with the same initial shock radius, typically $\RS(t=0)=100 \; \mathrm{km}$. In \S \ref{subsec:Lcrit_R_S}} we vary the initial shock radius, in order to demonstrate that it too affects the critical neutrino luminosity.

\begin{figure}[h]
\centering 
\fig{\includegraphics[width=\columnwidth]{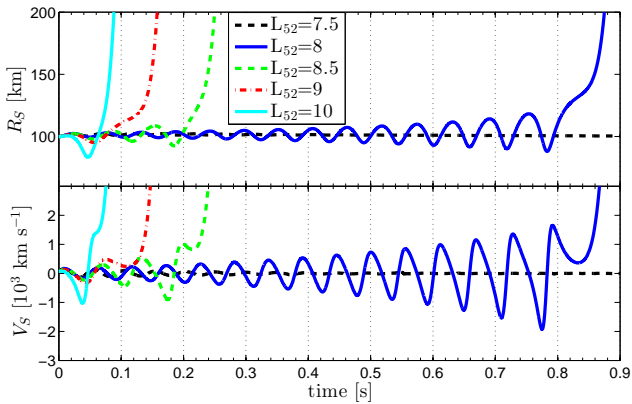}}
\caption{Simulations of shock dynamics [radius (top panel) and velocities (bottom panel)] for various luminosities [$L_{52}$ is the neutrino luminosity in units of $10^{52} \mathrm{erg  \ s^{-1}}$], both bellow (black dashed curve) and above (other curves) the critical luminosity. Fixed parameters in the simulations are: $|\dot{M}_0|=0.8 \mathrm{M_\odot\;s^{-1}}$, $\RS(t=0)=100 \; \mathrm{km}$ and $q_d=0$.} 
\label{fig:simulation}
\end{figure}

One of our goals is to identify the process of shock revival through the growth of oscillations, until the onset of runaway expansion, which -- following \citet{Fernandez2012} -- we define the onset of an oscillatory explosive flow. Examples of the time-dependent evolution of the shock radius and velocity for $M_P=1.3 M_\odot$, $\RP=40$km, $|\dot{M}_0|=0.8 \mathrm{M_\odot \ s^{-1}}$ and different neutrino luminosities are presented in Figure \ref{fig:simulation}. Dissociation energy losses at the shock were neglected $(q_d=0)$ in these simulations. All these simulations are initiated with the accretion shock set at $\RS(t=0)=100 \; \mathrm{km}$, with a stationary profile that corresponds to this radius. 

As seen in the figure, most simulations show some initial oscillations of the shock radius and velocity, which have typical periods of tens of milliseconds (see appendix \ref{app:t_osc} for the reason for this narrow range of periods). The distinction between an explosive and a nonexplosive case is clearly evident: when the neutrino luminosity is large enough, the oscillations grow in amplitude until eventually the shock velocity no longer reverts to a negative value, but rather continues to increase with a positive acceleration, culminating in a runaway expansion. The lowest neutrino luminosity which drives such a flow corresponds to the critical luminosity - in this case, about $L_{crit,52}\approx 8$, while for higher luminosities the onset of runaway expansion occurs earlier (after fewer oscillations); for $L_{52}= 10$, barely one oscillation is completed before the shock velocity growth becomes exponential (this can be considered a "non-oscillatory mode" \citep{Fernandez2012}). The curve for $L_{52}=7.5$ represents results typical of $L<L_{crit}$: small oscillations are damped and the flow settles back to stationary accretion.

\begin{figure} [h]
\centering
\fig{\includegraphics[width=\columnwidth]{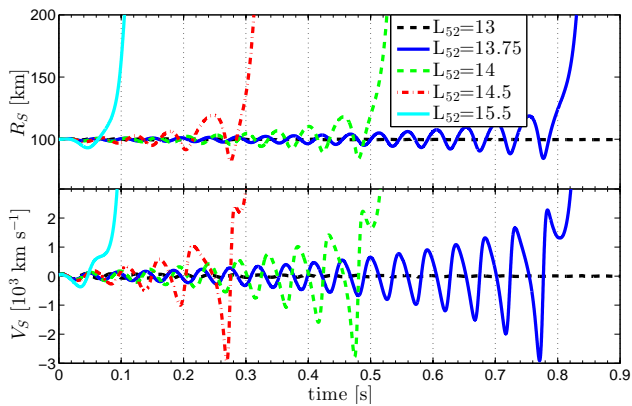}}
\caption{Shock dynamics as in Figure \ref{fig:simulation}, but including dissociation losses $q_d=q_{Fe}=8.5 \times 10^{18}  \mathrm{erg \ g^{-1}}$.}
\label{fig:simulation_dissociation_full}
\end{figure}

In Figure \ref{fig:simulation_dissociation_full} we show the results of simulations similar to those of Figure \ref{fig:simulation}, except that full dissociation losses at the shock are included with $q_d = q_{Fe}$. Evidently, the inclusion of dissociation losses does not qualitatively change the character of the flow; we use this feature in \S \ref{subsec:QS_approximation} when we apply the quasi-stationary model. The inclusion of dissociation energy losses does cause a shift in the dependence of the flow on the neutrino luminosity, requiring higher luminosities to achieve a similar flow. This is to be expected, of course, since increased luminosity is required to compensate for the energy loss rate  $|\dot{M}|q_d\approx1.4\times10^{52}\;\mathrm{erg \ s^{-1}}$. In practice, the critical luminosity is increased by a larger margin than $|\dot{M}|q_d$, to $L_{crit,52}\approx 13.75$, since dissociation also changes the compression factor at the shock, and hence the outer boundary conditions. 

\section{Shock Acceleration in a Phase-Space Diagram} \label{sec:shock_acceleration}

In this section we develop the quasi-stationary approximation to derive an approximate expression for the shock acceleration, $a_S$, given the shock radius, $\RS$ and velocity, $\VS$. This derivation allows us to identify regions in the $\RS-\VS$ plane in which both the velocity and acceleration remain invariably  positive, hence indicating in a phase-space fashion when the conditions are favorable for a runaway expansion.

First, consider the spherical (with respect to the center of the PNS) moment of inertia of the accretion layer between $\RP$ and $\RS$,
\begin{equation}\label{eq:I(t)}
I(t) \equiv \int_{\RP}^{\RS} r^2  \mathrm {dm} = \int_{\RP}^{\RS} 4\pi r^4 \rho \, \mathrm {dr}\; .
\end{equation}
Its first time derivative is
\begin{eqnarray} \label{eq:dIdt_initial}
\frac{\mathrm dI}{\mathrm d t} & = &
\int_{R_P}^{R_S} 4\pi r^4 \frac{\mathrm \partial \rho}{\mathrm  \partial t} dr + \left[ 4 \pi r^4 \rho \frac{dR}{dt} \right]^S_P \nonumber \\
& = & \int_{R_P}^{R_S} 4\pi r^4 \frac{\mathrm \partial \rho}{\mathrm  \partial t} dr+4 \pi R_S^4 \rho_S V_S \fin
\end{eqnarray}
The notation $[\cdots]^S_P$ stands for the difference between the expression in the immediate downstream of the shock and just outside the PNS, where the latter term is null in our model since we fix $dR_P/dt=0$.

We find that a general feature in the simulated flows is that to a very good approximation, the local mass accretion rate is uniform in the accretion layer, 
\begin{equation}
\dot{M}(r) \simeq \dot{M} \equiv 4 \pi r^2 \rho u \coma 
\end{equation}
corresponding to slow changes in the density profile. This reflects the highly subsonic nature of the downstream flow. Note that $\dot{M}$ is not identical to the incoming mass accretion rate, $\dot{M}_0$, since it is modified by the shock velocity. Indeed, the mass accretion rate just below the shock is then
\begin{equation} \label{eq:dotM_S}
\dot{M}_S=\dot{M}_0+4 \pi R_S^2 V_S \rho_0 (\beta-1) \fin
\end{equation}
For a radius-independent mass accretion rate, the first term on the right hand side of Equation (\ref{eq:dIdt_initial}) is approximately zero, and so
\begin{equation} \label{eq:dIdt_simple}
\frac{\mathrm dI}{\mathrm d t}\simeq4 \pi R_S^4 \rho_S V_S\;.
\end{equation}

Given that $dI/dt$ is now approximated as a function of $R_S$ and $V_S$, the second time derivative of the moment of inertia can be used to create an implicit relation for the shock acceleration,
\begin{equation} \label{eq:d2Idt2_derivative}
\left(\frac{\mathrm{d}^2 I}{\mathrm{d} t^2}\right) \simeq V_S \frac{\mathrm{\partial}}{\mathrm{\partial} R_S} \left( \frac{\mathrm{d} I}{\mathrm{d} t} \right)  + a_S  \frac{\mathrm{\partial}}{\mathrm{\partial} V_S} \left( \frac{\mathrm{d} I}{\mathrm{d} t} \right) \fin
\end{equation}
We emphasize that this is only an approximate equality due to the assumption of a uniform accretion rate in the entire accretion layer, resulting in $dI/dt$ being a function of $\RS$ and $\VS$ alone.

Equations (\ref{eq:dIdt_simple}-\ref{eq:d2Idt2_derivative}) relate bulk properties of the accretion layer, $dI/dt$ and $d^2I/dt^2$, to the quantities at the accretion shock. We find that this relationship is reproduced to a very high accuracy in our simulations. In principle, this relation can be used to calculate the shock acceleration for a given profile, allowing us to predict the oscillatory movement of the shock, and whether or not runaway is to be expected. Generally, however, $d^2I/dt^2$ cannot be derived without a dynamical simulation, for which a prediction of $a_S$ is redundant. Nevertheless, using some dedicated approximations, we derive in the following a quasi-stationary model, which does allow for the prediction of $a_S$, mapping it as a function of the instantaneous shock and flow properties.

\subsection{A Quasi-Stationary Approximation} \label{subsec:QS_approximation}

We modify the simple, stationary approximation by including a non-zero shock velocity when setting up the conditions at the shock and then solving equations (\ref{eq:RH_density}-\ref{eq:RH_energy}) as described above.
This approximation, which we refer to as "quasi-stationary", is applicable when the shock velocity is small with respect to the velocities in the accretion layer, which in turn are smaller than the typical sound speed in the flow. Note that the assumption of a uniform accretion rate implies that the velocity profile in the accretion layer adjusts quickly to changes in the shock radius, while a subsonic flow (also assumed in the stationary approximation) ensures that thermodynamic changes in the shock quickly advect throughout the shocked material, and influence the inner regions of the flow.

Quantitatively, these conditions can be assessed through the typical time scales in the flow, defined by the oscillation period, $t_{osc}$, 
\begin{equation}
t_{sc} \equiv \int_{R_p}^{R_S} \frac{\mathrm{d}r}{|c_s|}
\end{equation}
and
\begin{equation}
t_{adv} \equiv \int_{R_p}^{R_S} \frac{\mathrm{d}r}{|u|}\approx\frac{M_{env}}{\dot{M}} \coma 
\end{equation}
which are the sound crossing time and advection time respectively. In the second equality for the advection time, $M_{env}$ is the mass of the envelope, \ie the accretion layer. The second equality for $t_{adv}$ is exact only if the mass accretion rate is indeed uniform in the entire layer. To be precise, $t_{osc}$ is relevant when the flow goes through several oscillations before growing to runaway expansion (or damping out). Once the flow evolves to runaway expansion --- or alternatively, if the expansion is initially non-oscillatory --- the appropriate measure of the shock evolution should be its dynamical time, $t_S\equiv R_S/V_S$. 

In the parameter range of interest the post shock flow is very subsonic, with $t_{sc}$ being a few milliseconds and the shortest of the three timescales. The competition between the oscillation and advection time scales is more complicated, since the latter is very sensitive to the compression factor at the shock, and hence to dissociation losses. When neglecting dissociation losses we find that for luminosities close to or exceeding $L_{crit}$, 
the advection time scale is $t_{adv} \sim 10 \, \mathrm{ms}$, while the oscillation period is about $50-60\; \mathrm{ms}$. This typical value for the oscillation period can be quantitatively assessed with our quasi-stationary model; see Appendix \ref{app:t_osc}.

\begin{figure*}[t]
  \fig{\includegraphics[width=\textwidth]{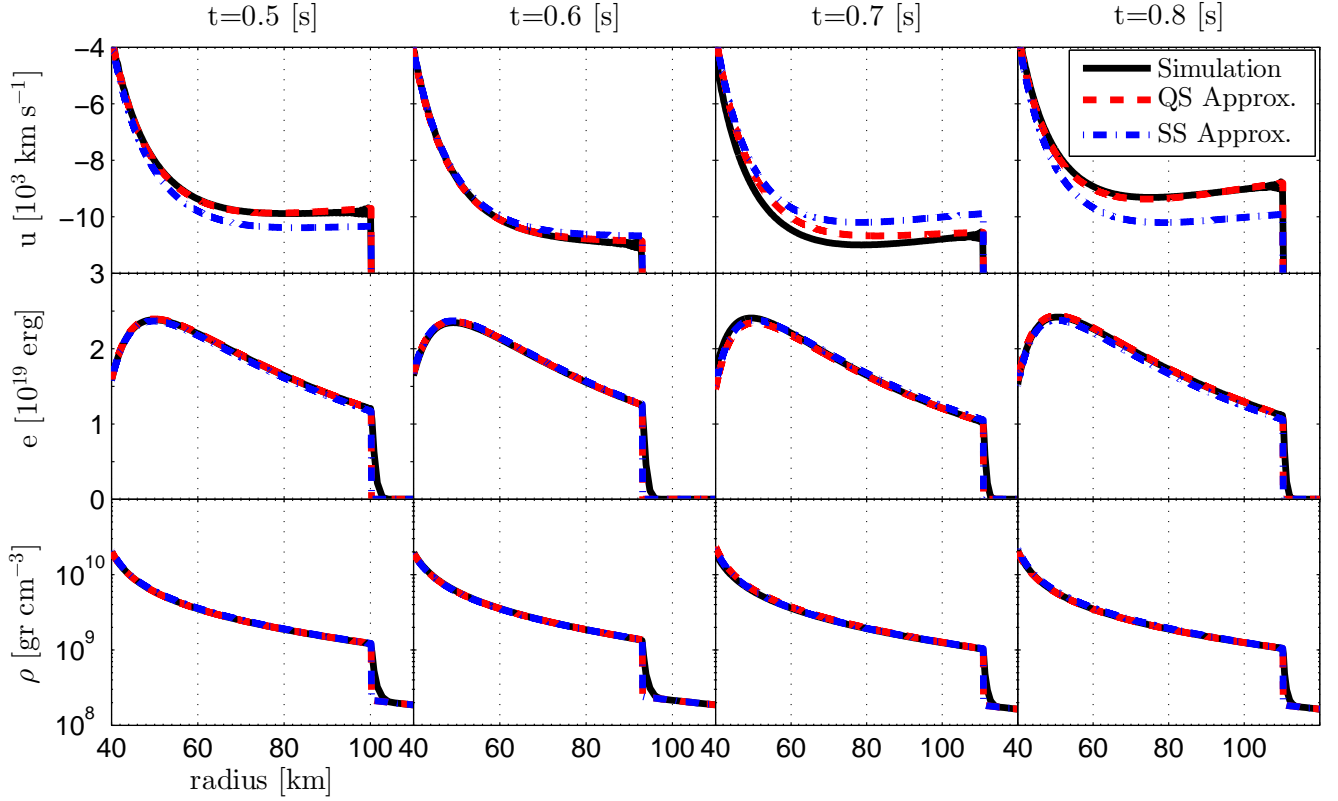}}
  \caption{Spherical symmetric simulation compared to quasi-static (QS) approximation and steady-state (SS) approximation, for $|\dot{M}_0|=0.8 \mathrm{M_\odot \ s^{-1}}$, $L_{52}=8$, $R_S(t=0)=100 $ km and $q_d=0$.
  Shown are the velocity (top), specific energy density (middle), and mass density (bottom), at four different times (see labels).
  The smoothing of the density profile at the simulated shock is a numerical effect, due to the use of artificial viscosity and the recipe for dissociation (equation \ref{eq:q_d_star}).}
\label{fig:compare_QS_to_SIM}
\end{figure*}

The hierarchy $t_{sc}<t_{adv}<t_{osc}$ implies that the quasi-stationary approximation should be applicable. We demonstrate this explicitly in Figure \ref{fig:compare_QS_to_SIM}, which compares four snapshots of the velocity, specific energy, and mass density profiles in the accretion layer, for the case $|\dot{M}_0|=0.8\mathrm{M_\odot \ s^{-1}}$, $L_{52}=8$ (the solid blue curve in Figure \ref{fig:simulation}), as found in the simulation and in the quasi-stationary approximation (using the instantaneous shock radius and velocity from the simulation). We also show the stationary profiles (found for the instantaneous shock radius but setting $\VS=0$) at every snapshot. Clearly, the quasi-stationary approximation offers a significant improvement over the stationary case, offering an almost exact fit not only in specific energy and density, but also in the velocity profile of the flow. Correspondingly, the quasi-stationary approximation we use below to analyze the shock acceleration is suitable for the regime of small dissociation losses - or gradual dissociation in the context of the microphysics of the shocked material.

The quasi-stationary approximation is not quite as successful in the opposite limit of full dissociation into free nucleons across the shock. For $q_d = q_{Fe}$, the compression factor at the shock is higher, generally between 10 and 20, which results in a low fluid velocity; for luminosities close to critical we find that $t_{adv}$ can reach 20 ms. On the other hand, $t_{osc}S$ is relatively insensitive to the level of dissociation, and so the quantitative deviations of the actual shock acceleration from that predicted by the quasi-stationary model become larger. Nonetheless,  we find that the qualitative behaviour of the flow and runaway expansion is similar even in the limit of full dissociation (since the oscillation period remains the largest time scale in the problem), and so we do assess that the quasi-stationary approximation is generally a useful starting point in the qualitative analysis of the transition to runaway expansion.

The advantage of the quasi-stationary approximation is that it allows us to estimate the second time derivative of the moment of inertia, $(d^2I/dt^2)_{QS}$, which has units of energy. The full derivation is given in Appendix \ref{app:d2Idt2}; here we quote the final result,
\begin{equation} \label{eq:d2I/dt2_QS}
\frac{1}{2} \frac{\mathrm d^2 I}{\mathrm d t^2}  \simeq \mathbb{E}_{QS}+W_{PNS}+W_B \coma
\end{equation}
where the three components on the right hand side are an effective energy, $\mathbb{E}_{QS}$, a work term associated with the PNS, $W_{PNS}$, and the energy advected across both boundaries, $W_B$. These are defined as follows. 

The effective energy, composed of kinetic, gravitational, and internal contributions, 
can be written in the form
\begin{equation}
\mathbb{E}_{QS}\approx \tilde{K}+\tilde{\Omega}+\tilde{U}+\tilde{B}_S
\approx \tilde{K}+\tilde{\Omega}+\tilde{U}\coma
\end{equation}
where the last expression can be estimated from the quasi-steady model. 
Here, we defined
\begin{equation} \label{eq:effective_K}
\tilde{K} \equiv  \int_{R_P}^{R_S} \left( \frac{3-\gamma}{\gamma} \right) \frac{1}{2} u^2 \mathrm{dm}\;,
\end{equation}
\begin{equation} \label{eq:effective_gravity}
\tilde{\Omega} \equiv  \int_{R_P}^{R_S} \left(\frac{3-2\gamma}{\gamma} \right)  \left( -\frac{G M_P}{r} \right) \mathrm{dm}\coma
\end{equation}
\begin{equation} \label{eq:effective_U}
\tilde{U} \equiv  \int_{R_P}^{R_S} \left( 3 \frac{\gamma-1}{\gamma}  \right)  \left[\frac{1}{|\dot{M}|}Q(r) \right] \mathrm{dm}\coma
\end{equation}
and 
\begin{equation}\label{eq:effective_B_app1}
\tilde{B}_S \equiv \int_{R_p}^{R_s} \left( 3 \frac{\gamma-1}{\gamma}  \right)  b_S(m) \, \mathrm{dm}\coma
\end{equation}
where $\gamma$ is the local adiabatic index, 
$Q$ is the rate of change of the total internal energy due to heating and cooling in a layer extending between $r$ and the shock,
\begin{equation} \label{eq:Q_r}
Q(r) \equiv 
\int_r^{R_s} \dot{q} \;\mathrm{dm} 
=
\int_r^{R_s} 4 \pi  \rho r^2 \dot{q} \;\mathrm{dr} \coma
\end{equation}
and $b_S$ is the Bernoulli function (specific energy),
\begin{equation}
b \equiv \frac{1}{2} u^2 + e + \frac{p}{\rho} - \frac{G M_p}{r} \coma
\end{equation}
evaluated for every mass element according to its value when crossing the shock. The non-adiabatic (heating and cooling) contribution to the energy is contained in the double integral $\tilde{U}$, which tracks the total gained non-adiabatic energy in the accretion layer.

The non-standard coefficients of the energy integrands in Eqs.~(\ref{eq:effective_K}--\ref{eq:effective_U}) arise from writing $\mathbb{E}_{QS}$ in a form that is susceptible to the quasi-steady approximation. In this approximation, terms associated with the instantaneous advection that is implied by assuming a steady state with a moving shock, are isolated in $\tilde{B}_s$ and subsequently neglected; see appendix \ref{app:d2Idt2}.

Next, there is work done by the PNS, arising from static and ram pressures at $R_P$. In general, static pressure dominates, so that $W_{PNS}$ can be approximated by (see Appendix \ref{app:P_P}):
\begin{equation} \label{eq:W_PNS}
W_{PNS} \simeq 4 \pi R_P^3 P_P\;.
\end{equation}
Finally, the energy advected across the boundaries is
\begin{align} \label{eq:W_S}
& W_B \simeq \alpha (G M_P)^{1/2} |\dot{M}_0 |  R_S^{1/2} \\
& +  \frac{|\dot{M}_0 |}{7 \alpha (G M_P)^{1/2}} \frac{d}{dt} \left\{ \left [(\beta-1)\left(\frac{R_P}{R_S}\right)^2+1\right ]
\frac{d}{dt} \left( R_S^{7/2} \right) \right\} \;, \nonumber
\end{align}
where the second term is the correction that arises due to the finite shock velocity. 

Substituting our result for $(d^2I/dt^2)_{QS}$ in Equation (\ref{eq:d2Idt2_derivative}), we arrive at a closed form expression for the shock acceleration in the quasi-stationary approximation:
\begin{equation} \label{eq:acceleration_QS}
a_S \approx \frac{ \mathbb{E}_{QS}  + W_{PNS}
 - |\dot{M}_0 R_S u_0| -V_S^2 \frac{\mathrm{\partial} \zeta}{\mathrm{\partial} R_S} }{\chi \zeta} \coma
\end{equation}
where
\begin{equation}\label{eq:zeta}
\zeta \equiv \frac{\left( \beta-1 \right)  \left( R_S^2-R_P^2 \right) |\dot{M}_0| } { 2|u_0|}
\end{equation}
and
\begin{equation}
\chi \equiv \frac{ 1 }{\beta -1}  \frac{ \mathrm{\partial} }{ \mathrm{\partial} V_S } \left[ V_S \left( \beta-1 \right) \right]\;.
\end{equation}
All the variables in equation (\ref{eq:acceleration_QS}) are determined by the set of external parameters of the flow (including $P_P$, which we fix at the beginning of the simulation), and the dynamical variables $\RS$ and $\VS$ (recall that $\beta=\beta(R_S,V_S,M_P,\dot{M}_0)$). 

\subsection{The $a_S(\RS,\VS)$ Phase Space} \label{subsec:phase_space}

\begin{figure*}[p]
\centering
\fig{
\includegraphics[width=.8\textwidth]{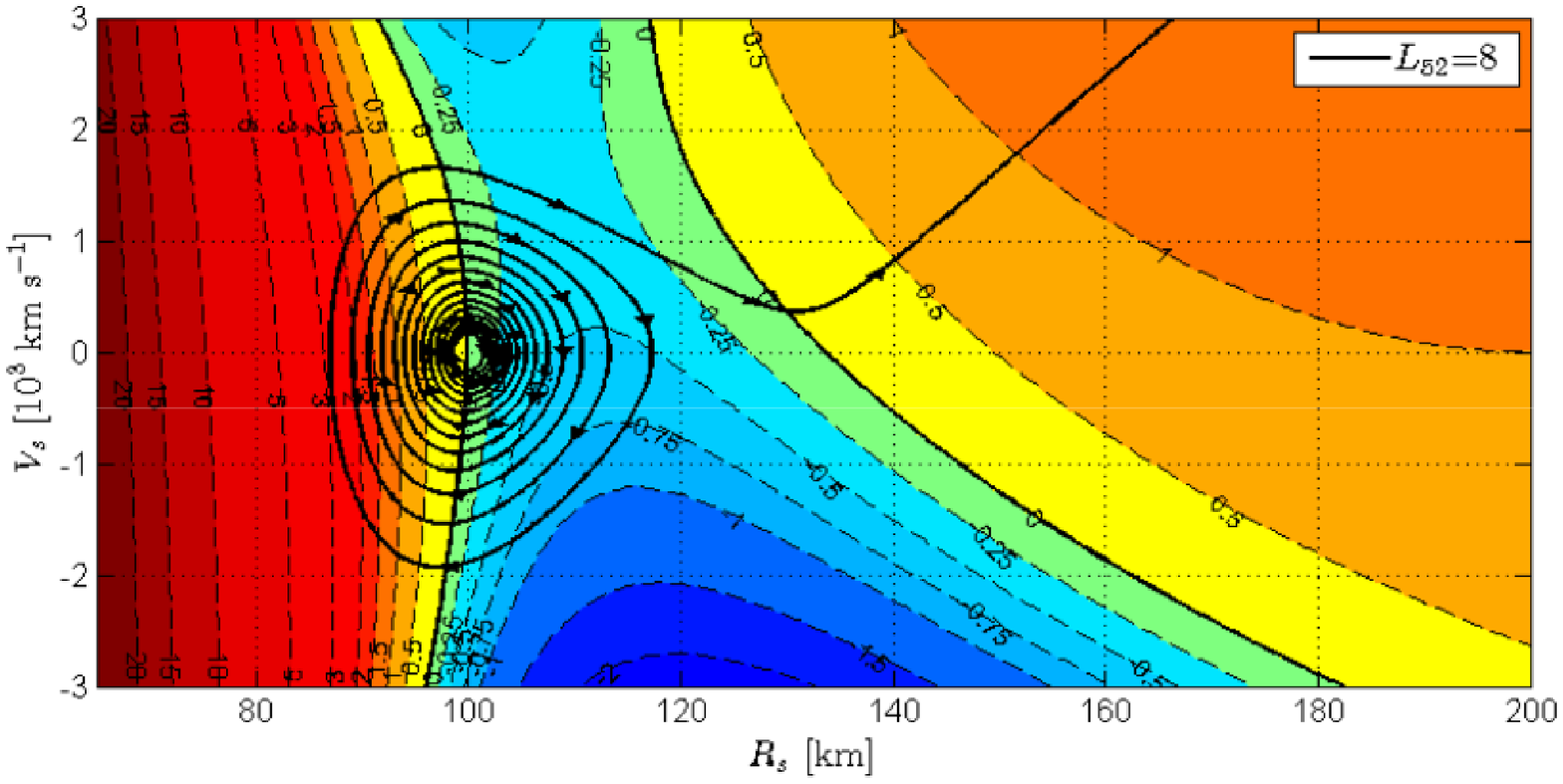}
\includegraphics[width=.8\textwidth]{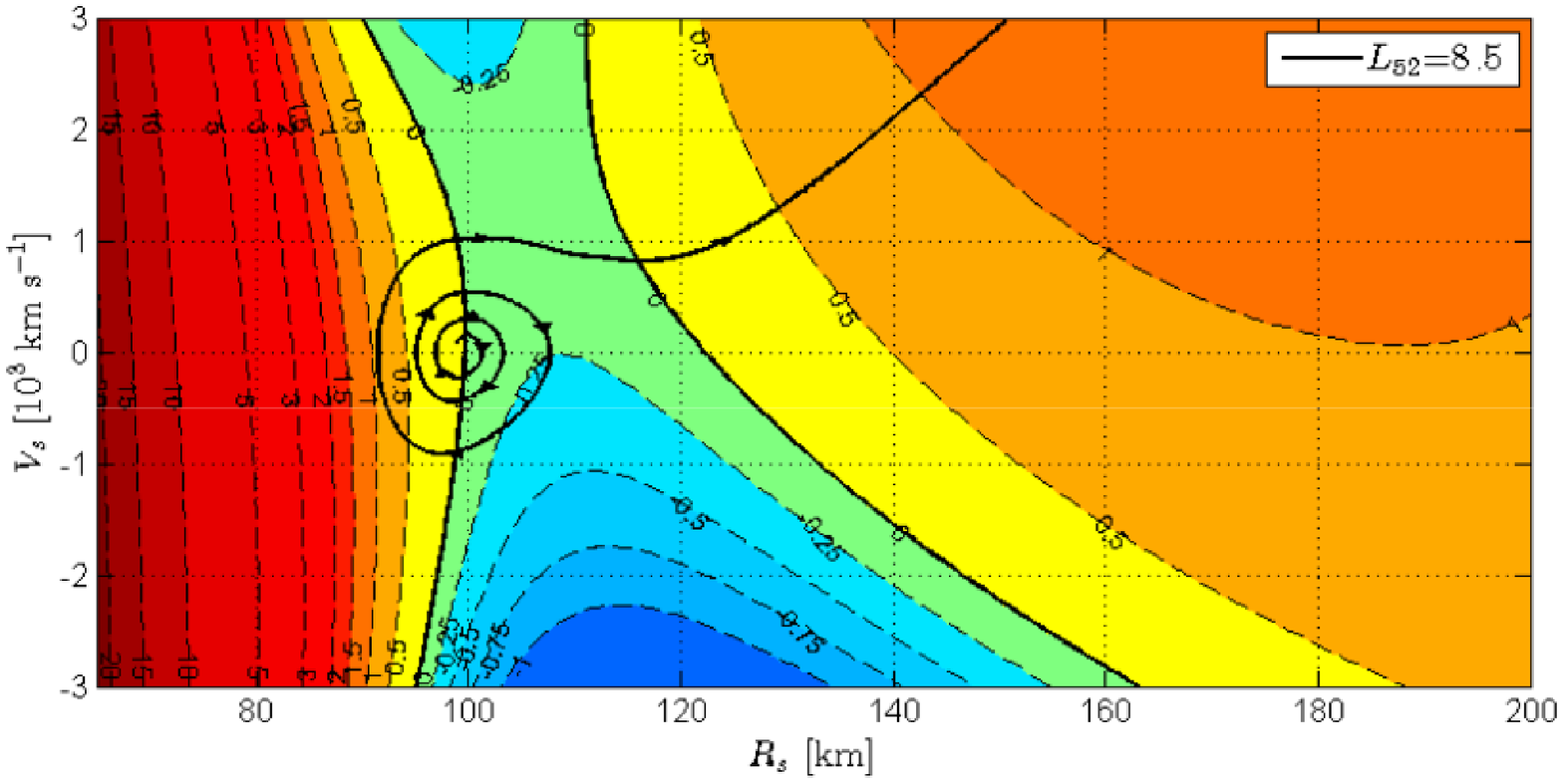}
\includegraphics[width=.8\textwidth]{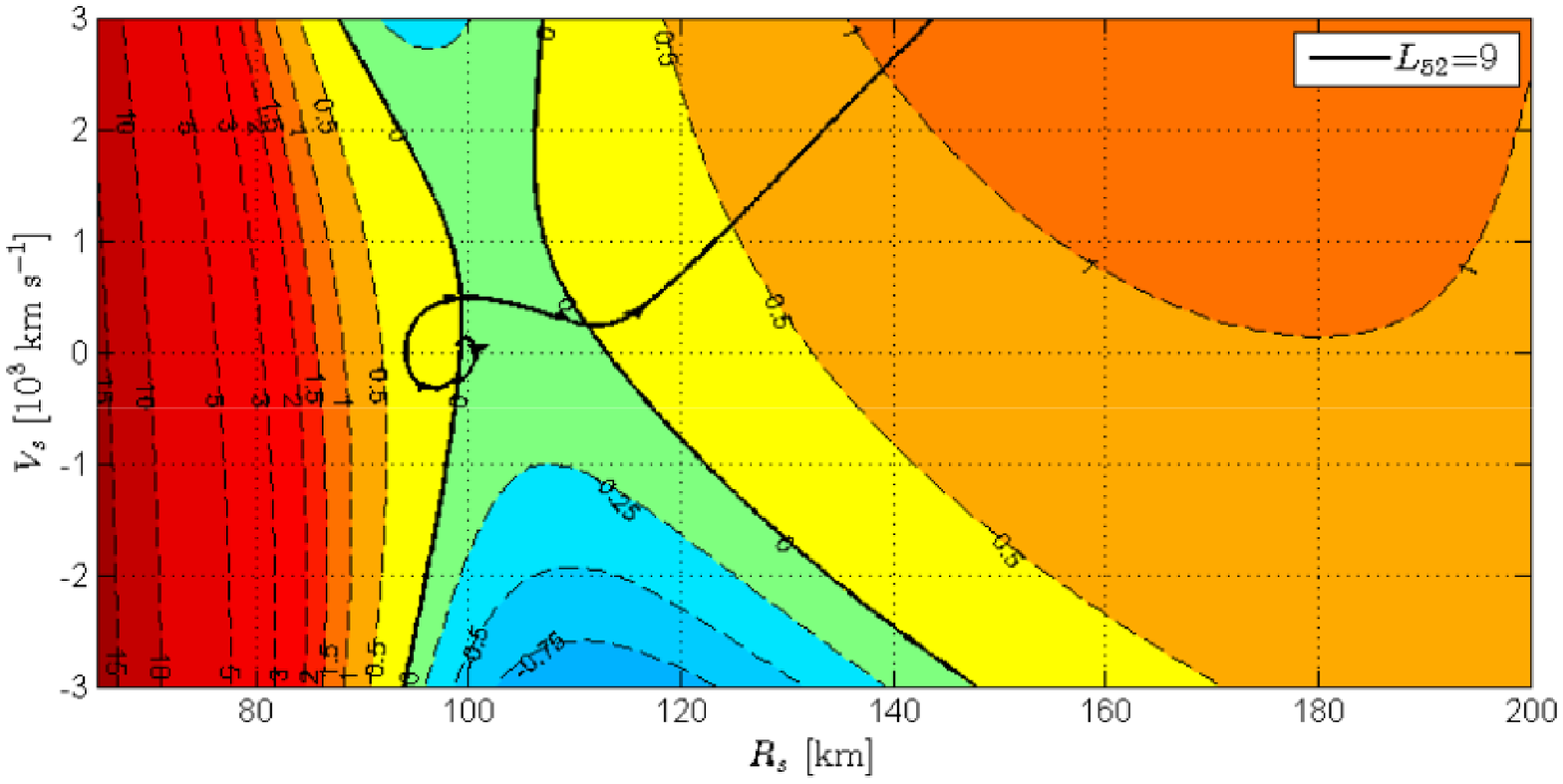}
}
\caption{Propagation of the shock in the $R_S-V_S$ phase-space for oscillatory ($L_{52}=8,8.5$) and marginally non-oscillatory ($L_{52}=9$) expansion (note that we define shocks that runaway after a single cycle as non-oscillatory). The quasi-stationary acceleration, calculated in equation (\ref{eq:acceleration_QS}), is plotted as contour lines in units of $10^6 \;\mathrm{km \ s^{-2}}$. Flow parameters are: $|\dot{M}_0|=0.8 \mathrm{M_\odot s^{-1}}$, $R_S(t=0)=100$ km, and $q_d=0$. Arrows indicate the direction in which the simulated shock evolves, and are marked every 25 ms.}
\label{fig:phase_space_dMdt0.8}
\end{figure*}

With the aid of the quasi-stationary approximation, we calculate the partition of the $\RS-\VS$ plane into regions of positive and negative acceleration of the shock. Given a set of external parameters, $\dot{M}_0$, $L$, $M_P$ and $\RP$, and the conditions of the initially stationary shock, determined by $\RS(t=0)$ and $\VS(t=0)=0$ (which also dictate the boundary conditions at the PNS), every point in the phase-space is calculated with the relevant quasi-stationary profile, yielding an estimate of the shock acceleration. A set of examples with $|\dot{M}_0|=0.8 \mathrm{M_\odot \ s^{-1}}$, $M_P=1.3 \mathrm{M_\odot }$, $\RP=40$ km and $\RS(t=0)=100$ km are shown in Figure \ref{fig:phase_space_dMdt0.8}. The figure shows the predicted values of $a_S$ in the $\RS-\VS$ plane for three neutrino luminosities which are sufficient to drive a runaway expansion (see Figure \ref{fig:simulation}). The actual evolutionary path of the shock radius and velocity, as found in each simulation, is superimposed on each of the plots.

The most prominent feature in Figure \ref{fig:phase_space_dMdt0.8} is the distinct separation of the $\RS-\VS$ plane into regions of positive and negative acceleration. At small and large shock radii the acceleration is positive, separated by a trough of negative acceleration. In the region of small shock radii, the acceleration is dominated by the effect of the PNS surface pressure, which is why in this region the dependence on the shock velocity is relatively weak, while the spatial gradient of the acceleration is quite steep. In contrast, for larger shock radii, a positive acceleration depends on a more complicated combination of the total energy in the flow, and on the boundary conditions at the shock itself, both of which depend on the shock velocity as well.

A key result is the fact that the actual evolutionary paths of the shocks in the $\RS-\VS$ plane follow the predicted acceleration of the quasi-stationary model very closely, oscillating between negative and positive velocities in accordance with the sign of the acceleration. We find that the approximation for the shock acceleration is qualitatively robust, but tends to overestimate its magnitude when compared to the full simulations (the term $B_S$ neglected above does tend to reduce $|a_S|$). Nonetheless, the contours of zero acceleration usually correspond very accurately to the points along the path where the shock acceleration in the simulations changes sign, especially for oscillating shocks. This result lends strong support to the validity of the quasi-stationary approach and the phase-space analysis. 

The transition to a runaway expansion occurs when the oscillations grow sufficiently large such that a velocity of order $1000\;\mathrm{km \ s^{-1}}$ pushes the shock through the negative acceleration trough into the region of positive acceleration on the right hand side of the phase space. Typically, the path passes close to the saddle point of $a_S(\RS,\VS)$ and arrives in the right-handed region of positive acceleration with a positive velocity, so the shock continues to expand. Hence, when the shock arrives in this region, the oscillations (if present) cease, and exponential expansion ensues. We hereon refer to the right curve (larger $\RS$) of zero acceleration as the "critical" $a_S=0$ curve. In contrast, the left $a_S=0$ curve (smaller $\RS$) corresponds to the change of sign of the shock acceleration in oscillatory motion, and does not necessarily signify a transition to runaway expansion. We refer to this curve as the "oscillatory" $a_S=0$ curve.

It is important to note that we do not find a case where the shock acceleration becomes negative at larger radii, to the right of the critical curve. The reason for this trend is that as the shock radius increases further, the competition between heating and other energetic effects maintains a positive acceleration (Equation \ref{eq:acceleration_QS}). Specifically, while $\tilde{\Omega}$ and ($-|\dot{M}_0 R_S u_0|$) become more negative with increasing shock radius, the effective heating term, $\tilde{U}$, becomes larger (more positive), and generally dominates the change in $a_S$ (we note that $W_{PNS}$ remains constant while $\tilde{K}$ and the last term in Equation (\ref{eq:acceleration_QS}) are essentially negligible). Hence, once the evolving shock crosses the critical $a_S$ from left to right, runaway expansion is inevitable. 

\begin{figure*}[p]
\centering
\fig{
\includegraphics[width=.8\textwidth]{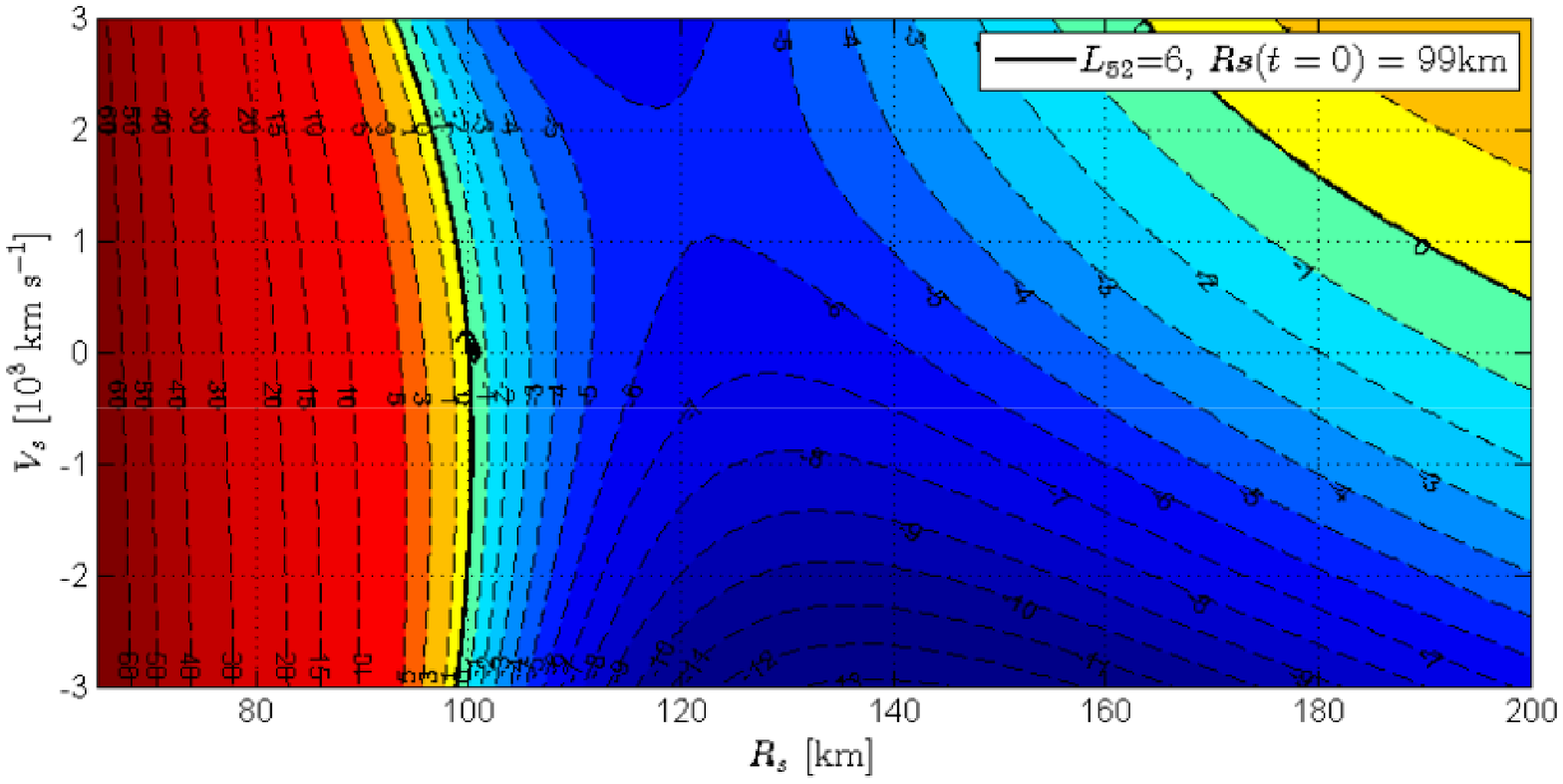}
\includegraphics[width=.8\textwidth]{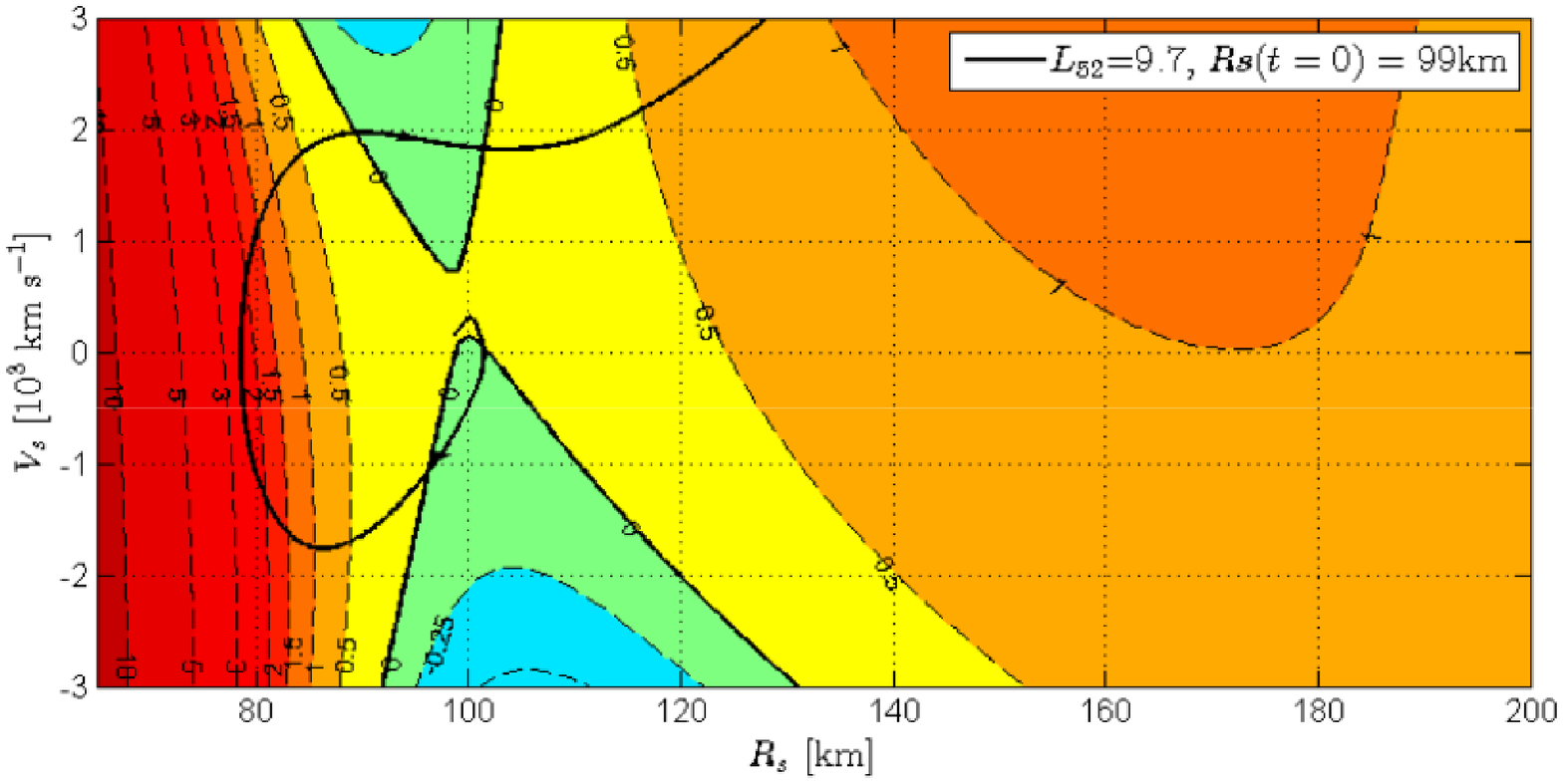}
\includegraphics[width=.8\textwidth]{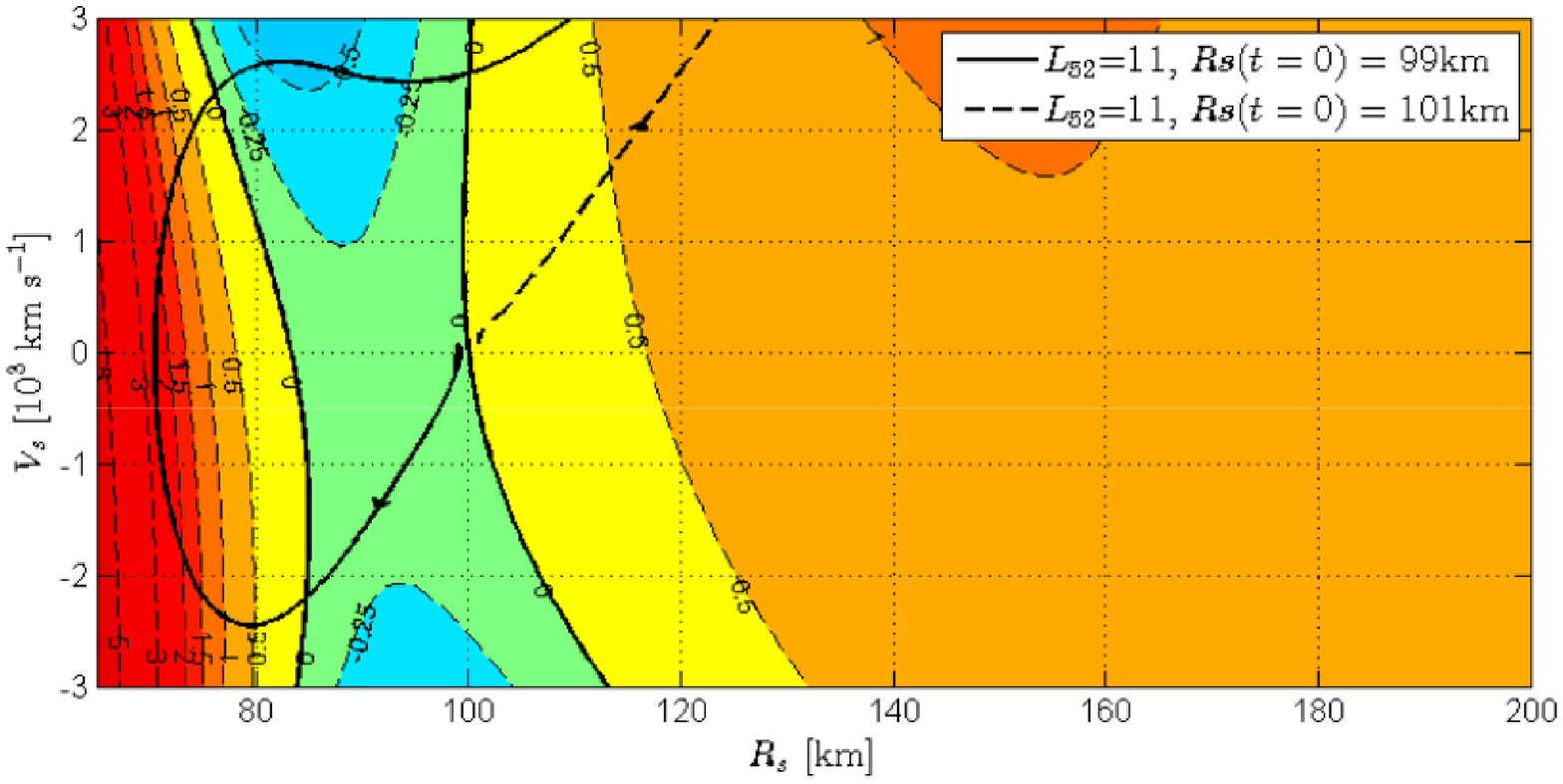}
}
\caption{Same as Figure \ref{fig:phase_space_dMdt0.8}, but for:
top panel  - $L_{52}=6$ (damped oscillations); middle panel - $L_{52}=9.7$ (near the point where the $a_s=0$ curves intersect); and lower panel: $L_{52}=11$ - an initial profile on the critical curve with exponential runaway, either directly (dashed) or after one oscillation (solid).}
\label{fig:phase_space_dMdt0.8_part2}
\end{figure*}

The magnitude of the neutrino luminosity has a substantial effect on the acceleration phase-space, as is readily seen when comparing the three plots in Figure \ref{fig:phase_space_dMdt0.8}. Clearly, as we raise $L$ the negative acceleration trough becomes narrower and shallower, and so the shock requires fewer oscillations to reach a point in the $\RS-\VS$ plane which will carry it over to the positive acceleration region. To further elucidate this point, we show in Figure \ref{fig:phase_space_dMdt0.8_part2} the phase space and evolutionary paths for a subcritical luminosity of $L_{52}=6$ (see Figure \ref{fig:simulation}), for a transitional luminosity $L_{52}= 9.7$, and for a high luminosity of $L_{52}=11$. In the subcritical case, small perturbations are damped due to local stability (see \S \ref{subsec:paths_to_explosion}), eventually converging onto a stationary solution, whereas larger perturbations are bounded due the wide and deep negative acceleration area in the phase-space. 
As the luminosity increases, criticality is reached as the damping turns into anti-damping, and the shock continues to expand through a series of oscillations. Increasing the luminosity further narrows the negative acceleration trough until eventually the two $a_S=0$ curves intersect. At even higher luminosities the curves detach again, and this detachment reflects a qualitative change in the explosive evolution. Now the initial profile lies on the critical $a_S=0$ curve rather than on the oscillatory one (recall that we dictate initial profiles with $V_S(R_S=100\mathrm{\ km})=0$ and $a_S(R_S=100\mathrm{\ km})=0$). Such an initial profile will necessarily end in runaway expansion if the initial perturbation drives it into larger shock radii, but we find that it may also reach runaway expansion after one oscillation, if the perturbation is in the opposite, radially negative, direction. We conclude that the intersection of the $a_S=0$ curves corresponds to a transition between oscillatory and direct runaway expansion.

\subsection{Paths to Runaway Expansion} \label{subsec:paths_to_explosion}

\begin{figure*}[t]
  \centering
\fig{\includegraphics[width=\textwidth]{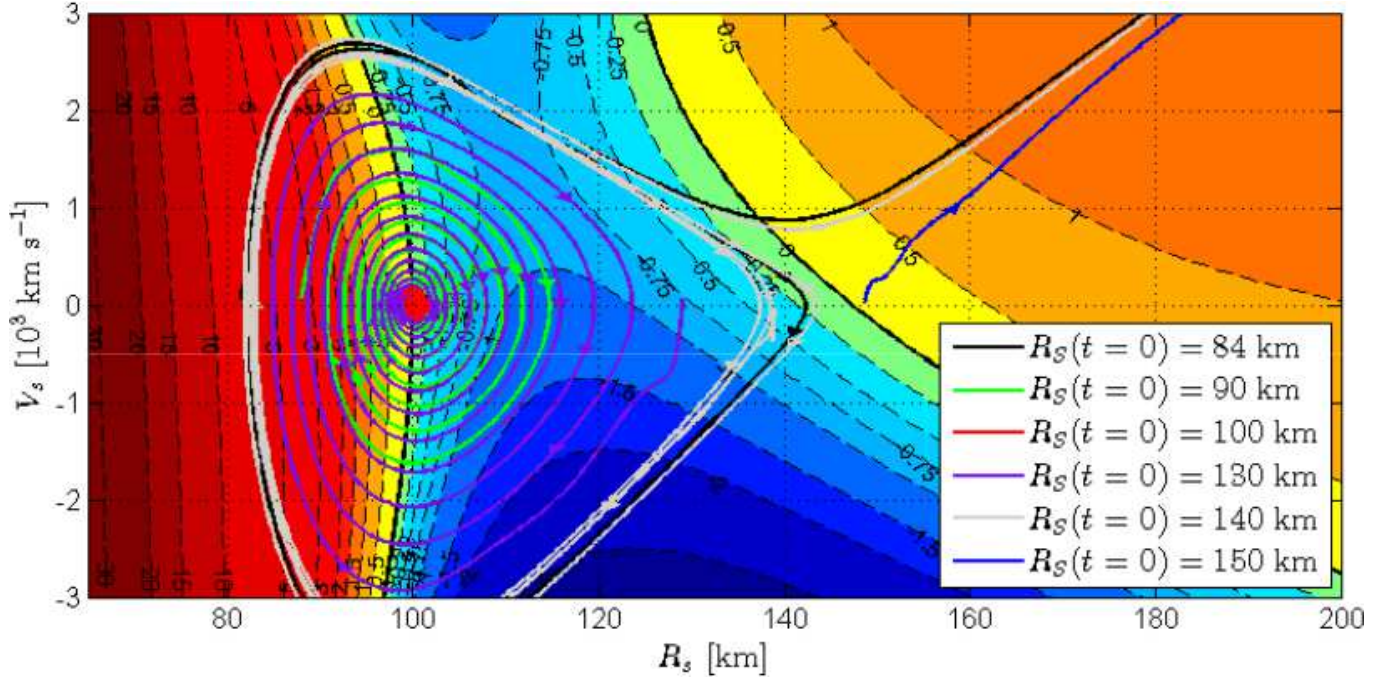}}
  \caption{Evolution of initially static ($V_S=0$) shocks with different initial radii in the phase-space diagram. In all simulations $L_{\nu_e,52}=7.5$, $|\dot{M}_0|=0.8 \mathrm{M_\odot \ s^{-1}}$ and $q_d=0$. The PNS pressure (and the overall phase space) is calculated for a stable accretion shock at $R_{s,0}=100$ km. The Flow is unstable for initial radii outside of $84 \mathrm{km} \le R_S(t=0) \le 150 \ \mathrm{km}$}
\label{fig:change_Rs0}
\end{figure*}

The phase space interpretation leads us to the following conclusion: even when all external parameters - $\{\dot{M}_0$, $M_P$, $R_P$, $P_P$ and $L\}$ - are set, both the initial shock radius and velocity determine whether or not the flow will become unstable and reach runaway expansion. We demonstrate this by means of an example in Figure \ref{fig:change_Rs0}, in which we repeat the simulations in the case of $L_{52}=7.5$, including the same pressure at the PNS, but start with static $V_S(t=0)=0$ shocks at different shock radii. In order to show the evolutionary paths corresponding to these different radii superimposed on a single $a_S(R_S,V_S)$ phase space map, we do not adjust the boundary PNS pressure at according to each initial shock radius ($P_P$ is set for $R_S(t=0)=100$ km). Consequently, most profiles have a non-zero initial shock acceleration, and evolve to cover a wide range of shock radii and velocities. The plot shows that there exist small and large initial radii for which the flow results in a runaway expansion, while in an intermediate regime of shock radii it does not. Conversely, we can consider any combination of $\{\RS,\VS\}$ as an initial profile for the simulation. There is a finite region in the $\RS-\VS$ plane (its border lying between the paths labeled $R_S=130$ km and $R_S=140$ km) for which oscillations are eventually damped, and the flow settles on the stationary solution. The physical reason for this limited range of $\{\RS,\VS\}$ for which runaway expansion is denied (in this particular setup) is that at sufficiently small radii the pressure close to the PNS creates a large positive shock acceleration, which can drive the shock all the way beyond the critical $a_S=0$ curve. Hence, runaway expansion will occur either if the flow is initiated in this region, or if is initiated in combinations of $\{R_S,V_S\}$ which eventually evolve so that the shock penetrates deeply into the left handed side region of positive acceleration. Of course, the critical $a_S=0$ curve serves as an additional limit: a profile which is initiated in that region will accelerate exponentially if the initial shock velocity is positive (or even only slightly negative).

\nopagebreak

In the general case of arbitrary initial $\{R_S,V_S\}$ the outcome of the evolution can be found by calculating the entire evolutionary sequence (directly or by following it after mapping the appropriate phase space). Moreover, when the problem is restricted to initially stationary profiles
$(V_S(t=0)=0,\; a_S(t=0)=0)$, the situation is greatly clarified and we find that the outcome of the evolution can be assessed by the initial signs of the partial derivatives $\partial a_S / \partial R_S$ and  $\partial a_S / \partial V_S$. This can be seen by noting that here, perturbations are governed by 
\begin{equation}
\label{eq:LinearOscillator}
\ddot{R}_S=a_S=\frac{\partial a_S}{\partial R_S}(R_S-R_{S,0})+\frac{\partial a_S}{\partial V_S}\dot{R}_S \coma
\end{equation}
which is simply the damped (or anti-damped) linear oscillator equation. This suggests the following behavior.

\begin{itemize}
\item {An initially stationary profile with $\partial a_S/\partial R_S>0$ necessarily corresponds to a point on the critical $a_S=0$ curve. As is the case in the lower panel of Figure \ref{fig:phase_space_dMdt0.8_part2}, we find that such a profile will invariably evolve to a runaway expansion in a non-oscillatory fashion. In essence, this class of initially stationary profiles is inherently unstable.  }
\item{An initially stationary profile with $\partial a_S/\partial R_S<0$ necessarily corresponds to a $\{R_S,V_S=0\}$ point on the oscillatory $a_S=0$ curve. Whether an initial perturbation diverges or damps out can be gauged locally by the sign of
$\partial a_S / \partial V_S$. 
If $\partial a_S / \partial V_S<0$, we expect a stable scenario: oscillations will tend to damp out, and the flow will settle onto the stationary solution. 
If this derivative is positive, runaway expansion is expected, but its nature is more subtle. In a purely local perturbation theory, we would expect that a further distinction will arise from the sign of  
$\Delta\equiv (\partial a_S/\partial V_S)^2+4\partial a_S/\partial R_S$. If $\Delta<0$, the shock should evolve through a series of increasing ("anti-damped") oscillations, finally resulting in runaway expansion. On the other hand, $\Delta>0$, along with $\partial a_S / \partial V_S>0$ leads to non-oscillatory, runaway expansion (even though $\partial a_S/\partial R_S<0$).}
\end{itemize}

\begin{figure*}[t]
  \fig{\includegraphics[width=\textwidth]{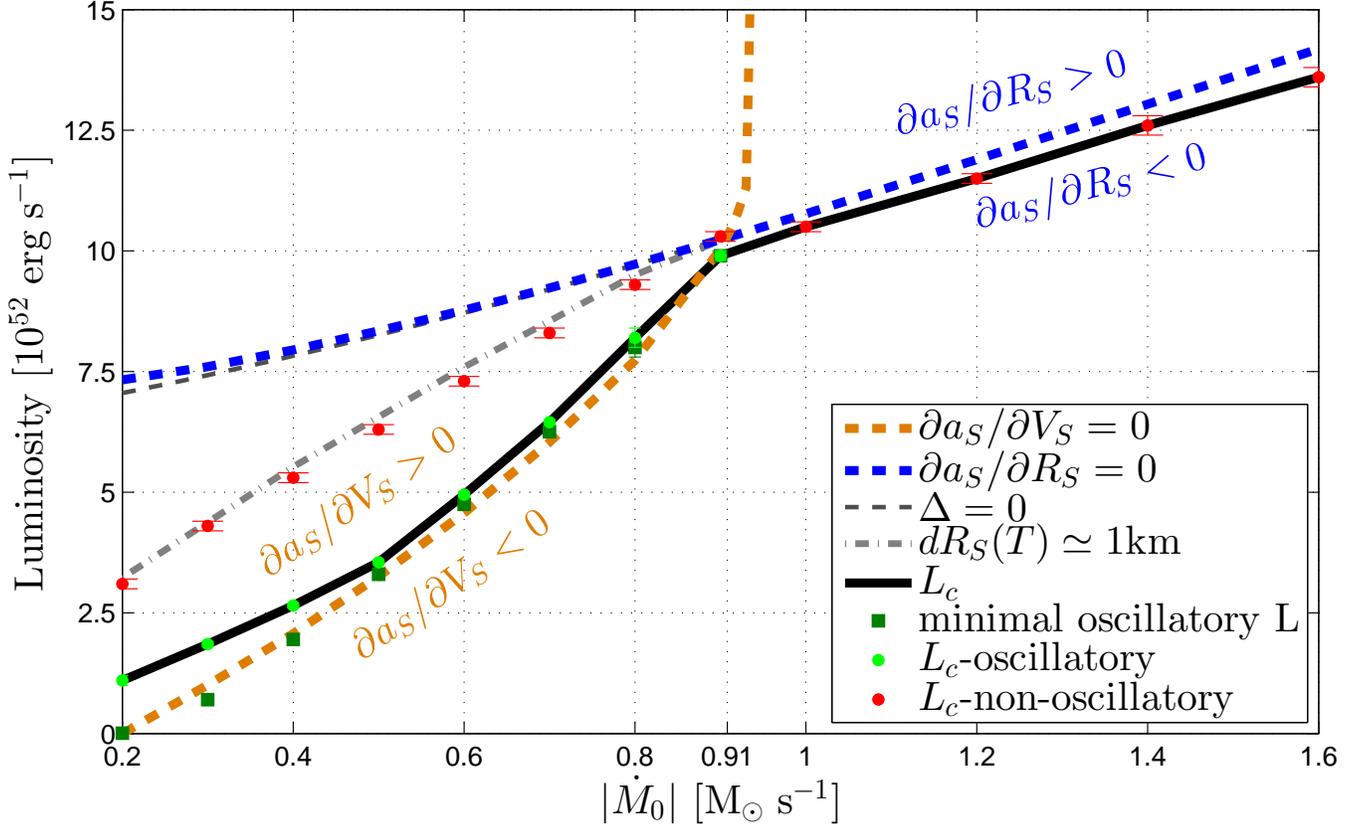}}
\caption{The critical luminosity as a function of the external mass accretion rate. 
Single points summarize a survey of simulations and show the minimal neutrino luminosities for stable oscillations (dark green squares), oscillatory runaway (green dots), and non-oscillatory runaway (red dots). The black line represents the combined critical luminosity. These data are based on our highest resolution, and are accurate to within $5\%$; error bars, which are too small to see at low $|\dot{M}_0|$, represent the $1\sigma$ distribution of the numerical survey.
Also shown are curves calculated with the quasi-stationary approximation: $\partial a_S/\partial V_S=0$ (orange dashed line; $\partial a_S/\partial V_S>0$ above this curve), $\partial a_S/\partial R_S=0$ (blue dashed line; $\partial a_S/\partial R_S>0$ above this curve), $\Delta=0$ (gray dashed; $\Delta>0$ above this curve), and when perturbations are expected to grow to $\simeq 1 \mathrm{km}$ in a single oscillation period (gray dash-dot).  See text for details.}
\label{fig:Lcrit_Mdot_daSdRS_daSdVS}
\end{figure*}

We test this insight in Figure \ref{fig:Lcrit_Mdot_daSdRS_daSdVS}. In the figure we show the results of a survey of simulations, presenting the critical luminosities as a function of mass accretion rate for $M_P=1.3\;M_\sun$ and $R_P=40\;$km. Each simulation is initiated with a small perturbation around a stationary profile at $R_S=100\;$km. The survey allows to distinguish between stable configurations and runaway expansion, and also to distinguish whether runaway occurs through an oscillatory or a  non-oscillatory path. We define runaway expansion as non-oscillatory when there is no more than one cycle where the shock radius drops below its initial value; for example, the $L_{52}=9$ case in Figure \ref{fig:simulation} is an example of a marginally non-oscillatory runaway expansion. The figure shows the critical luminosities for oscillatory and non-oscillatory runaway expansion, $L_{c,osc}$, and $L_{c,dir}$, respectively. These are compared to curves which show the loci of $\partial a_S/\partial R_S=0$, $\partial a_S/\partial V_S=0$, and $\Delta=0$ as calculated with the quasi-stationary approximation (i.e., finding all combinations of $L$ and $|\dot{M}_0|$ for which an initial stationary profile with $R_S(t=0)=100$ km would yield these derivatives).

As seen in the figure, there is a good qualitative consistency between the simulated critical luminosities and the theoretical curves, evaluated by the shock acceleration in the quasi-stationary approximation. The  $\partial a_S/\partial R_S=0$ curve is especially indicative of the threshold for runaway expansion through the non-oscillatory path at higher mass accretion rates, for which $\partial a_S/\partial V_S<0$ for all luminosities. Interestingly, the $\partial a_S/\partial V_S=0$ curve strongly depends on the accretion rate, requiring exceedingly higher $L$ for increasing $|\dot{M}_0|$, eventually diverging at $|\dot{M}_0|\simeq \dot{M}_c \equiv 0.91\, M_\sun\;{\mathrm s^{-1}}$. On the other hand, the $\partial a_S/\partial R_S=0$ curve has a rather weak dependence on the mass accretion rate, so for $|\dot{M}_0|>\dot{M}_c$ only one class of initial stationary profiles exists: this simply reflects the fact that for the higher accretion rates, initial profiles with $R_S(t=0)=100$ km that explode lie on the critical $a_S=0$ curve, rather than the oscillatory one. As is indeed confirmed in the simulations, runaway expansion in this regime can occur only through a non-oscillatory evolution. We note that at very high mass accretion rates, the $a_S$ phase space analysis becomes increasingly inaccurate due to a large contribution from the Bernoulli function, which cannot be neglected in highly non-oscillatory expansions, when the shock dynamical time becomes comparable to the advection time (see Appendix \ref{app:d2Idt2}).   

\begin{figure}[h]
  \fig{\includegraphics[width=\columnwidth]{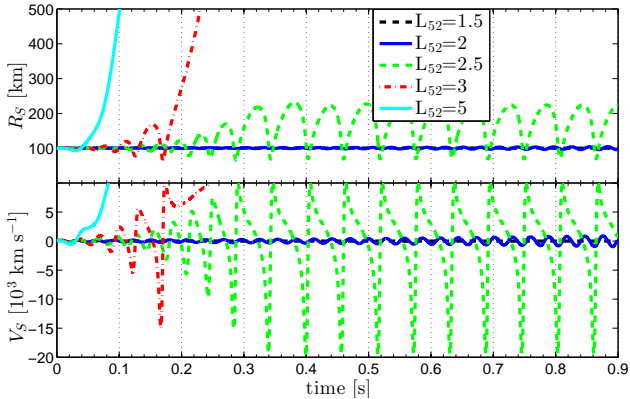}}
  \caption{Same as Figure \ref{fig:simulation}, but for $|\dot{M}_0|=0.4 \mathrm{M_\odot\;s^{-1}}$. Note that increasing the luminosity changes the evolution from damped oscillations to stable oscillations, and only then to oscillatory runaway, and finally to non-oscillatory runaway.}
\label{fig:RSvst_Mdot0.4}
\end{figure}

The parameter region where $\partial a_S/\partial R_S<0$ displays a somewhat more complicated behavior, although also qualitatively consistent with the quasi-stationary approximation. First, when $\partial a_S/\partial V_S<0$ (low luminosities), runaway expansion is inherently denied: initial perturbations are damped, and the system relaxes to the stationary profile, as expected. The $\partial a_S/\partial V_S=0$ curve, above which oscillations are expected to anti-damp and lead to runaway expansion, indeed lies very close to the critical luminosity curve $L_{c,osc}$, and hence provides a very good estimate for the critical condition for this type of evolution. The two curves do not quite coincide, and for most $\dot{M}_0$ there is a small range of luminosities for which $\partial a_S/\partial V_S>0$ but runaway expansion does not yet occur. We find that in this range oscillations commence, but they to tend to settle on a constant or semi-constant (increasing very slowly) amplitude. An example of such an evolution is presented in Figure \ref{fig:RSvst_Mdot0.4}, which shows the time-dependent radius of the shock for $|\dot{M}_0|=0.4M_\sun\;{\mathrm s^{-1}}$ and different neutrino luminosities. There is a general resemblance to Figure \ref{fig:simulation}, but the case of $L_{52}=2.5$ is unique: initially the oscillations grow, but then settle on an approximately constant amplitude. Presumably, the existence of such an evolutionary path is due to the non-local nature of the $a_S(V_S)$ relation (in other words, the oscillator in Equation (\ref{eq:LinearOscillator}) reaches amplitudes where it becomes non-linear).

Figure \ref{fig:Lcrit_Mdot_daSdRS_daSdVS} also shows the limiting luminosities for which there is a transition from oscillatory to non-oscillatory runaway expansion in the $\partial a_S/\partial V_S>0$ region. According to linear theory, this curve should coincide with the $\Delta=0$ curve (which is very close to the $\partial a_S/\partial R_S=0$ curve, and so is of little importance for the phase space analysis). However, in this particular aspect we are strongly constrained by numerical resolution, which prevents us from initiating the simulation with truly infinitesimal perturbations. In general, it is impractical to initiate the simulations with perturbations that are smaller in amplitude than about two hundred meters, and the eventual dynamics of the shock depend on the changes of $\partial a_S/\partial V_S$ and $\partial a_S/\partial R_S$ over this scale. In the simulations, a perturbation of this magnitude can grow in radius by about one kilometer during a single oscillation, which is significant in terms of the gradients of the acceleration. We show this by plotting in Figure \ref{fig:Lcrit_Mdot_daSdRS_daSdVS} a curve which corresponds to a growth of the initial perturbation to one kilometer in a single oscillation \footnote{In a linear oscillator, the amplitude of a perturbation will grow by a factor of $\exp[2\pi (\partial a_S/\partial V_S)/\sqrt{|\Delta|}]$, so a one kilometer amplitude will be reached in a single oscillation from the initial perturbation when the growth factor is four to five.}. 
The quantitative fit to the critical non-oscillatory luminosity suggests that finite resolution leads us to identify non-oscillatory runaway when linear analysis breaks down over a 1 km scale - obviously below the $\Delta=0$ curve. 
In any case, this transitional luminosity is less important than the critical luminosities, which separate exploding and non-exploding profiles.

\subsection{Dependence on the Initial Shock Radius} \label{subsec:Lcrit_R_S}

\begin{figure}[h]
\centering
\fig{\includegraphics[width=\columnwidth]{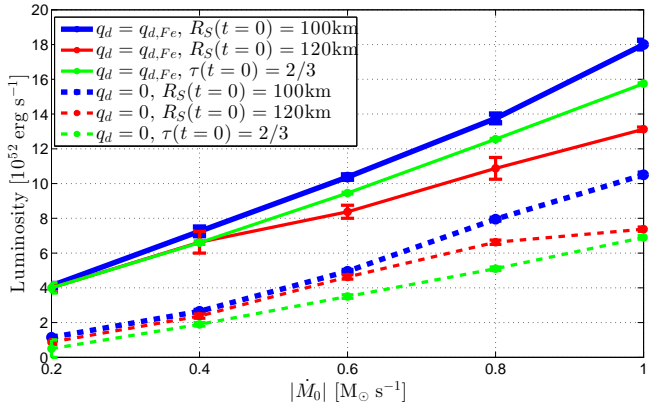}}
\caption{The critical luminosity as a function of the mass accretion rate for different initial values of the shock radius. Blue and red lines correspond to a fixed, $R_S(t=0)$ of 100 and 120 km, respectively, and the green line for an initial shock radius that corresponds to an initial optical depth of $\tau=2/3$ in the accretion layer. Dashed curves: simulations with no dissociation across the shock, $q_d=0$; solid curves: simulations with a full dissociation across the shock, $q_d=q_{Fe}$.}
\label{fig:Lcrits}
\end{figure}

So far we have set the initial shock radius arbitrarily at $R_S(t=0)=100$ km. Our phase space analysis suggests that there should be a non-trivial dependence of the evolution, and therefore of the critical luminosity on this radius, even when the set of parameters $\dot{M}_0, M_P$, and $R_P$ is fixed. In Figure \ref{fig:Lcrits} we confirm this expectation for the case of initial stationary profiles, where we compare the critical luminosity as a function of the external accretion rate, $L_{crit}(\dot{M}_0)$, when $M_P=1.3\;M_\odot$, $R_P=40\;$km and the initial profiles were calculated for $R_S(t=0)=100\;$km and $R_S(t=0)=120\;$km. We also compare the results with the critical luminosity found when the initial conditions are determined by requiring a $\tau=2/3$ neutrino optical depth through the accretion layer, so that $R_S(t=0)$ is constrained for each simulation separately; for all but the lowest $|\dot{M}_0|$, this condition forces a larger initial shock radius, with $R_S(t=0)=150-160\;$km. It is evident from the figure that a dependence on $R_S(t=0)$ does indeed exist (weak at small $\dot{M}$, becoming more pronounced for larger accretion rates). Our results are consistent with the conclusions of \citet{Fernandez2012}, and are distinct from the stationary model, which generates a single physical solution when constrained by an additional condition.

Also shown in figure \ref{fig:Lcrits} is a similar comparison of sets of critical luminosities found in simulations which included full dissociation across the shock, $q_d = 8.5 \times 10^{18} \ \mathrm{erg \ g^{-1}}$.  The qualitative dependence upon $\RS(t=0)$ is retained when full dissociation is included, while all the critical luminosities are shifted upwards, (as expected). Notably, in this case the shock radius corresponding to $\tau=2/3$ generally lies between 100 and 120 km, and so the critical luminosity curve for a constant optical depth is now second from the bottom, instead of first . Once again, the general trend of the results suggests that our principal conclusions can be carried over to the realistic case when dissociation is accounted for across the shock and later during accretion. 

\begin{figure}[h]
\fig{\includegraphics[width=\columnwidth]{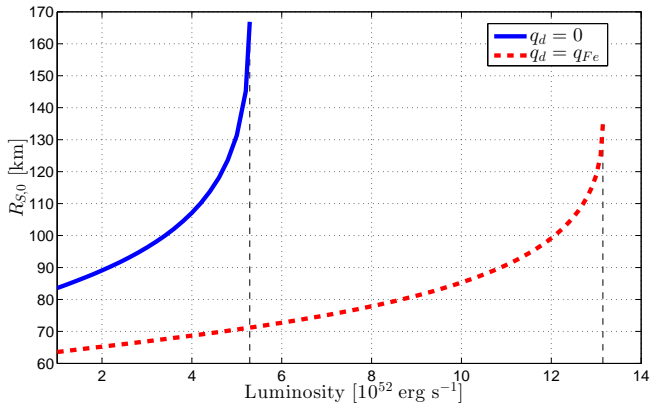}}
\caption{The initial radius of stationary profiles as a function of the neutrino luminosity when constrained by a total neutrino optical depth of the accretion layer of $\tau=2/3$. The curves correspond to models with no dissociation across the shock (solid blue curve) and full ($q_d=q_{Fe}$) dissociation (dashed red curve). The curves are terminated at a luminosity for which no solution can be found (the \citet{burrows_goshy93} limit). In these models $R_P=40\;$km and $|\dot{M}_0|=0.8 \mathrm{M_\odot\;s^{-1}}$.}
\label{fig:R_S0_L_BG}
\end{figure}

The \citet{burrows_goshy93} critical luminosity, $L_{c,BG}(\dot{M})$, is determined by the absence of a stationary solution. In cases where a stationary $\tau=2/3$ solution exists, but its instability leads to runaway expansion, the critical luminosity in simulations, $L_c(\dot{M})$, must be lower than $L_{c,BG}(\dot{M})$. A more subtle issue here is that when using the $\tau=2/3$ constraint, $L_{c,BG}$ is generally only a mild overestimate of $L_c$ even though the issue of stability is ignored \citep{Fernandez2012}. We qualitatively relate this to the phase space analysis by noticing that in stationary models, as the neutrino luminosity is increased towards $L_{c,BG}$, the initial shock radius changes significantly - see Figure \ref{fig:R_S0_L_BG}. As a result, the phase space structure of $a_S(R_S,V_S)$ changes in a substantial manner at luminosities close to (but lower than) $L_{c,BG}$, transforming the system to a non-stable one. 
Thus, the critical luminosities corresponding to the disappearance of a solution and to its instability lie in close proximity.  

\section{A Timescale Condition for Runaway Expansion?} \label{sec:explosion_condition}

Our phase space interpretation, combined with the quasi-stationary approximation, enabled us to identify the initial signs of the partial derivatives of the shock acceleration as a criterion for the prospects of an explosion when starting from a stationary profile. 
We emphasize that even though they are expressed by quantities at the shock, these are global criteria, in the sense that they depend on the properties of the entire accretion layer. In this section we use the phase space analysis and the quasi-stationary approximation to evaluate a commonly discussed timescale criterion for the onset of an explosion, comparing heating and advection in the so-called "gain region" of the flow \citep{murphy_burrows08,pejcha_thompson12,Fernandez2012,Ottal2013}.

The specific neutrino heating scales as the inverse of the radial position of the mass element in the accretion layer, while the specific neutrino cooling scales as the sixth power of the element's temperature (recall Equations (\ref{eq:heating}-\ref{eq:cooling})). As a result, at smaller radii where the temperatures are higher, cooling dominates over heating, $\dot{q}(r)<0$, whereas closer to the shock heating dominates and $\dot{q}(r)>0$. Such a structure is also found in supernovae simulations, including studies which use more accurate modeling than our simplified equations \citep{Fernandez2012,Ottal2013}. Correspondingly, it is common practice to identify the radial position in the flow where $\dot{q}=0$ as the "gain radius", $R_G$, and refer to the region extending between this radius and the shock as the "gain region".

The argument concerning the timescales is that if in the gain region the heating time,
\begin{equation}\label{eq:t_heatG}
t_{heat,G}= \frac{E_G}{\int_{R_G}^{R_S} \dot{q} \mathrm{dm}}\;,
\end{equation}
is smaller than the corresponding advection time in this region,
\begin{equation}\label{eq:t_advG}
t_{adv,G}= \int_{R_G}^{R_S} \frac{\mathrm{dr}}{u} \approx \frac{M_G}{|\dot{M}|}\;,
\end{equation}
then the flow should evolve towards runaway expansion. In these equations, $E_G$ and $M_G$ are the total internal energy of matter and the total mass in the gain region, respectively, and the second equality for $t_{adv}$ is exact if the mass accretion rate is uniform. The criterion $t_{heat,G}<t_{adv,G}$ reflects the expectation that if the material can heat significantly before it passes through the gain region, steady state accretion could be disturbed. In the following, we demonstrate that the phase space structure of $a_S$ can be approximately related to this ratio of timescales. 

\begin{figure*} [t]
\centering
\fig{\includegraphics[width=15cm,height=15cm,keepaspectratio]{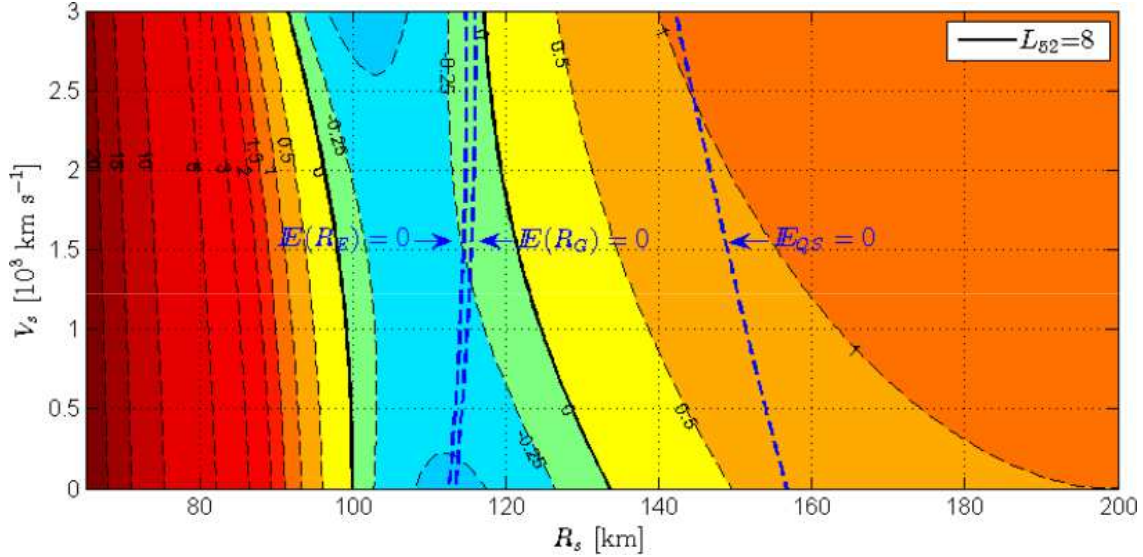}}
  \caption{Phase-space diagram (see Figure \ref{fig:phase_space_dMdt0.8}) with constant $\mathbb{E}=0$ lines. See text for details.}
\label{fig:phase_space_E_lined}
\end{figure*}

First, note that in Equation (\ref{eq:acceleration_QS}) the prefactor $\zeta$ is invariably positive, and $\chi$ is positive for positive shock velocities ($V_S(\beta-1)$ tends to increase with shock velocity). Hence, the sign of $a_S$ is determined by the expression inside the brackets in this equation. This expression includes the effective energy $\mathbb{E}_{QS} \approx \tilde{K}+\tilde{U}+\tilde{\Omega}$ which characterizes the entire accretion layer, and three additional terms which include contributions from the PNS and the shock. In Appendix \ref{app:P_P} we show that in the region of interest, the PNS term $W_{PNS}\simeq 4\pi R^3_P P_P$ dominates over the advection terms, and so we can generally claim that $\mathbb{E}_{QS}>0$ along with $V_S>0$ provide a {\it sufficient} condition for runaway expansion. We show this explicitly in Figure \ref{fig:phase_space_E_lined}, in which we plot again the shock acceleration in the $R_S-V_S$ plane, and superimpose upon it the $\mathbb{E}_{QS}>0$ curve. As expected, the $\mathbb{E}_{QS}>0$ curve lies in the runaway expansion region to the right of the critical $a_S=0$ curve, clarifying that a positive effective energy is indeed a sufficient - but not necessary - condition. 

\begin{figure} [h]
  \fig{\includegraphics[width=\columnwidth]{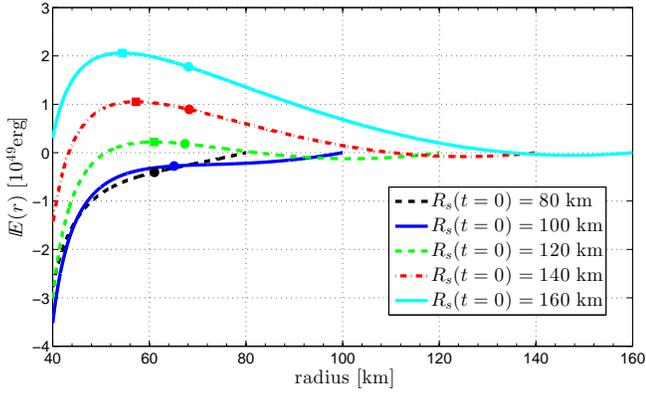}}
  \caption{The total effective energy $\mathbb{E}(r)$ for $|\dot{M}_0|=0.8 \mathrm{M_\odot \ s^{-1}}$, $L_{52}=8$ and $q_d=0$ for different $R_S(t=0)$. The locations of $R_G$ and $R_E$ are marked by circles and squares, respectively. See text for definitions of these quantities.}
\label{fig:E_total_vs_R}
\end{figure}

The total energy $\mathbb{E}_{QS}$ is the result of an integration over the entire accretion layer. Conversely, we can define a radius-dependent effective energy, $\mathbb{E}(r)$, corresponding to the integrals (\ref{eq:effective_K}-\ref{eq:Q_r}) taken from $r$ to $R_S$. In general, we expect $\mathbb{E}(r)>\mathbb{E}(R_P)\equiv \mathbb{E}_{QS}$, since matter close to the PNS experiences the strongest cooling and gravitational potential. As a result, there can exist combinations of $\RS$ and $\VS$ for which the profile has $\mathbb{E}(r)>0$ for some range of radii $R_P<r<R_S$, while $\mathbb{E}_{QS}<0$. Numerically, we find that for most initial configurations that eventually lead to runaway expansion, the function $\mathbb{E}(r)$ has a maximum at some inner radius in the accretion layer, which we denote as $R_E$. This radius is always close to, but slightly lower than $R_G$: just below the gain radius the internal energy still increases inward due to compression ($\partial \tilde{U}/\partial r<0)$, and this effect offsets the net cooling by neutrinos. We demonstrate the locations of $R_E$ and $R_G$ in Figure \ref{fig:E_total_vs_R}, which shows $\mathbb{E}(r)$ in the accretion layer calculated for various initial shock radii and a given set of external parameters ($L_{52}=8$, $|\dot{M}_0|=0.8 \ M_\sun \ s^{-1},\; M_P=1.3\ M_\sun$, and $q_d=0$). 

Repeating the process of identifying and calculating $R_E$ and $R_G$ as a function of $R_S$ and $V_S$ for a given set of external conditions, we find the loci of all the points in the $R_S-V_S$ plane for which $\mathbb{E}(R_G)=0$ and $\mathbb{E}(R_E)=0$. The two corresponding curves are also shown in Figure  \ref{fig:phase_space_E_lined}, and can be seen to lie very close to each other, and are both to the left of the critical $a_S=0$ curve. Our conclusion is that $\mathbb{E}(R_E)>0$ and $\mathbb{E}(R_G)>0$ are both thresholds that must be crossed when the flow evolves to runaway expansion, but they are not sufficient conditions conditions to ensure that runaway will occur. Indeed, for a wide range of parameters, crossing the $\mathbb{E}(R_G)=0$ curve does not ensure runaway expansion.

Finally, the quasi-stationary approximation can be used to relate the condition $t_{adv,G}/t_{heat,G}>1$ to the condition $\mathbb{E}(R_G)>0$.
With some manipulation of Equations (\ref{eq:p_div_rho}-\ref{eq:energy_QS}), we derive that
\begin{equation}
\mathbb{E}(R_G) \approx K_G + U_G + \left( 2 \hat{\gamma}_{E_G} -3 \right) E_G
\end{equation}
with $K_G$, $E_G$ and $U_G$ being the total kinetic energy, total internal energy and energy-gain between $R_G$ and $R_S$, respectively (the latter is the gain region equivalent to equation \ref{eq:energy_gain_U}),
and $\hat{\gamma}_{E_G}$ is the mean adiabatic index of the material in the gain region, weighted by the internal energy:
\begin{equation}
\hat{\gamma}_{E_G}=\frac{1}{E_G} \int_{R_G}^{R_S} \gamma e \mathrm{dm}\;.
\end{equation}
Since $E_G$ is positive by construction, the condition for $\mathbb{E}(R_G)>0$ becomes:
\begin{equation}\label{eq:bbE_G>0_1}
\frac{K_G}{E_G} + \frac{1}{E_G}\int_{R_G}^{R_S} \frac{1}{|\dot{M}|} Q(r) \mathrm{dm} > 3- 2\hat{\gamma}_{E_G}\;,
\end{equation}
where $Q(r)$ was defined in Equation (\ref{eq:Q_r}). Since $\dot{q} \ge 0$ in the gain region, $Q(r)$ monotonically decreases with the radius, and $Q(R_S)=0$. The total non-adiabatic integral over the entire gain region can therefore be approximated by:
\begin{equation}
\int_{R_G}^{R_S} Q(r) \mathrm{dm} = \eta Q(R_G) M_G
\end{equation}
where $\eta$ is a numerical factor reflecting the gradient of $Q(r)$ ($\eta=1/2$ for a linear function). We find that for the typical configurations in Figure \ref{fig:phase_space_dMdt0.8}, $\eta \sim 0.6$ and $\hat{\gamma}_{E_G} \sim 1.4$.

\begin{figure} [h]
\centering
\fig{\includegraphics[width=\columnwidth]{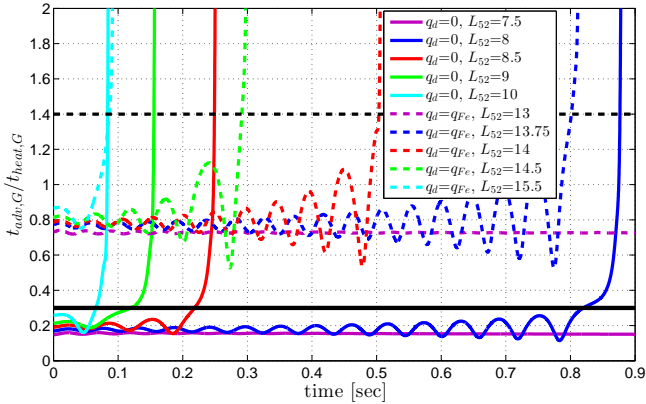}}
\caption{The ratio of the advection timescale to heating timescale as a function of time in simulations with $|\dot{M}_0|=0.8 \mathrm{M_\odot \ s^{-1}}$, and different neutrino luminosities $L$, in units of $ 10^{52} \mathrm{erg \ s^{-1}}$, $R_{s,0}=100$ km. Solid lines with $q_d=0$ and dashed lines with $q_d=8.5 \times 10^{18} \mathrm{erg \ g^{-1}}$. The ratios should be compared to the threshold of $t_{adv,G}/t_{heat,G}=0.3$, predicted by the quasi-stationary approximation for $q_d=0$ (black solid line) and $t_{adv,G}/t_{heat,G}=1.4$, predicted for for $q_d=0$.}
\label{fig:tadv_theat_qd0}
\end{figure}

Finally, since the accretion layer typically involves small velocities, the first term in Equation (\ref{eq:bbE_G>0_1}) is negligible, and so the condition for $\mathbb{E}(R_G)>0$ can be summarized as
\begin{equation}
\frac{\int_{R_G}^{R_S} \dot{q} \mathrm{dm}}{E_G} \frac{M_G}{|\dot{M}|} >\frac{1}{\eta}\left(3- 2\hat{\gamma}_{E_g}\right)
\end{equation}
or (using Equations (\ref{eq:t_heatG}) and (\ref{eq:t_advG}))
\begin{equation}\label{eq:tadv_theat_G}
\frac{t_{adv,G}}{t_{heat,G}}>\frac{1}{\eta}\left(3-2\hat{\gamma}_{E_G}\right).
\end{equation}
For the typical values of $\eta=0.6$, ${\gamma}_{E_G}=1.4$ and zero dissociation, the expression on the right hand side is about 0.3. A comparison of this result to actual simulations is presented in Figure \ref{fig:tadv_theat_qd0}. Indeed, runaway expansion is seen to commence in conjunction with the ratio of timescales in the gain region exceeding this threshold.

We note that dissociation losses can be included in the estimate for the required value of $t_{adv_G}/t_{heat,G}$ by correcting the non-adiabatic integrand to $Q(r)-q_d$. While this approximation is somewhat crude (since the quasi-stationary approximation is less appropriate for large dissociation losses), it does provide some quantitative estimate of the effect. The actual derivation then yields a correction of $+(q_d M_G )/(E_G \eta)$ to the right hand of Equation (\ref{eq:tadv_theat_G}), demonstrating that a larger dissociation energy requires a larger timescale ratio, since heating must compensate for the lost energy. With the aid of the quasi-stationary approximation we predict that in the case of full dissociation the necessary value of $t_{adv_G}/t_{heat,G}$ is about 1.4. Again, this value appears to be consistent with the simulations, as demonstrated in Figure \ref{fig:tadv_theat_qd0}.

Finally, we point out that the value of $t_{adv,G}/t_{heat,G}$ increases to the from left to right (increasing radii) in the $R_S-V_S$ plane, so a ratio of unity should in general not be far from the critical $a_S=0$ curve. However, the two are by no means identical criteria, and we find no distinct relation between $t_{adv,G}/t_{heat,G}$ and any specific feature in the phase space. Moreover, the value of unity does not appear to hold any particular significance: it occurs well into runaway in the absence of dissociation, while it is an insufficient condition (occurs during the oscillations prior to runaway) when full dissociation is assumed.

We summarize our analysis of the ratio between advection and heating time scales with two claims:
\begin{itemize}
\item{Strictly speaking, $t_{adv,G}/t_{heat,G}>1$ is not a special condition for an explosion. Rather, it is a value that apparently lies in the path of a shock that is on its way to a runaway expansion, and that it occurs close to (before or after) the transition to a positive shock acceleration\footnote{A similar argument was raised independently by \citet{Murphy_Dolence2015}, just as this manuscript was being completed.}.}
\item{We also do not identify the gain region as being in any way unique in terms of a necessary condition for runaway expansion. The curves in Figure \ref{fig:phase_space_E_lined} imply that $\mathbb{E}(R_E)>0$, or some other alternative curve, could be used just as well a necessary, but not sufficient, condition. Clearly, runaway expansion requires significant heating, but evolution towards a runaway expansion is a quality of the entire accretion layer, rather than just of the gain region.}
\end{itemize}

\section{Summary and Discussion}\label{sec:conclusions}

In this work we investigated the driving of a stalled accretion shock into runaway expansion in the context of the neutrino heating mechanism in core-collapse supernovae. We used spherically symmetric simulations in conjunction with analytic derivations, and examined the evolutionary paths the system may take towards runaway expansion, if it occurs. Our numerical results included a parameter survey of the neutrino luminosity and the incoming accretion rate, as well as of the shock parameters, namely, its instantaneous radius and velocity. Our main conclusions are as follows.

\begin{enumerate}
\item The instantaneous shock acceleration, $a_S$, depends on both the shock radius, $R_S$, and velocity, $V_S$. We examine the shock acceleration in the $R_S-V_S$ plane, and show that it is generally divided into regions of positive and negative shock acceleration (see Figures \ref{fig:phase_space_dMdt0.8} and \ref{fig:phase_space_dMdt0.8_part2}). We expanded the commonly used stationary approximation, by including a non-zero shock velocity and solving the resulting profile in the accretion layer with modified boundary conditions (\S \ref{subsec:QS_approximation}). This quasi-stationary approximation developed here allows to map this plane in advance of a time-dependent simulation, and nicely predicts the evolution of a model. The quasi-stationary approximation is especially accurate when the shock dynamical time, $t_S=R_S/V_S$, and oscillation period, $t_{osc}$, are much longer than the advection time through the accretion layer, which is the case if the immediate dissociation energy losses across the shock are not too large. We also use the derivation to estimate the oscillation period, and show why it is typically $25$--$50\;$ms, very weakly dependent on the mass accretion rate (see Appendix \ref{app:t_osc}). This result is consistent with simulations. 

\item Generally, the $\RS-\VS$ phase space shows positive shock acceleration at small and large shock radii, with an intermediate trough of negative acceleration. The trough is bounded by two $a_S=0$ curves. The one corresponding to smaller radii is an "oscillatory" $a_S=0$ curve: if the shock moves from it inward, positive acceleration due to the high pressure close to the PNS can cause the shock to bounce back and oscillate. Such oscillations either damp out or grow, depending on the parameters of the flow. In contrast, the $a_S=0$ curve at large radii is "critical", since crossing it with a positive velocity, leads to a region of positive acceleration, so runaway expansion is guaranteed (i.e., profiles which correspond to this curve are inherently unstable). The magnitude of the neutrino luminosity comes into play through quantitative aspects of this phase space structure. For low luminosities, the trough of negative acceleration is both deep and wide, and when the initial shock radius and velocity correspond to this trough, this shock is generally stable and oscillations are damped. When the neutrino luminosity is high, the trough of negative acceleration becomes shallow and narrow, and larger regions of it correspond to profiles which are unstable, so the shock eventually evolves beyond the critical $a_S=0$ curve and towards runaway expansion.

\item For arbitrary combinations of $R_S$ and $V_S$ serving as initial conditions, the evolution of a model can be tracked in the phase space estimated by the quasi-stationary approximation. Our analysis demonstrates that the shock velocity must be accounted for when assessing whether the evolution will end in runaway expansion or not. 

\item In the special case of initial stationary profiles ($V_S=0,\;a_S=0$), we find that the partial derivatives of the shock acceleration offer a satisfactory indication regarding the outcome. If $\partial a_S/\partial R_S>0$ and $\partial a_S/\partial V_S<0$ the initial profile lies on the critical $a_S=0$ curve, and the profile will evolve to runaway expansion in a non-oscillatory fashion. In contrast, if $\partial a_S/\partial R_S<0$ the initial profile lies on the oscillatory $a_S=0$ curve and its evolution depends on the sign of $\partial a_S/\partial V_S$. If this derivative is also negative, the system is stable and any small initial perturbation will damp out and settle to the stationary profile. On the other hand, if $\partial a_S/\partial V_S>0$, oscillations will tend to grow unstably (be "anti-damped"), leading to runaway expansion through an oscillatory path. This analysis bares out very well in the simulations, in the sense that we can predict the critical luminosity as a function of mass accretion rate, $L_{c}(\dot{M}_0)$, for both oscillatory and non-oscillatory explosions; see Figure \ref{fig:Lcrit_Mdot_daSdRS_daSdVS}. Quantitatively, $\partial a_S/\partial V_S=0$ is a very good indicator for finding the critical luminosity for oscillatory explosions, and $\partial a_S/\partial R_S=0$ does well in predicting the critical luminosity for non-oscillatory explosions. We note that we slightly underestimate $L_{c}(\dot{M}_0)$ for the oscillatory mode, because there is a small range of luminosities which generate stable oscillations. 

\item Both modes of runaway expansion are available for low mass accretion rates. In this case, oscillatory runaway with $\partial a_S/\partial V_S>0$ is met at lower luminosities than $\partial a_S/\partial R_S>0$, so the actual critical luminosity is due to anti-damped oscillations. We find that the luminosity at which $\partial a_S/\partial V_S=0$, $L_{c,osc}$, is strongly dependent on the mass accretion rate, and there exist some finite rate $\dot{M}_c$ for which $L_{c,osc}$ diverges. At larger accretion rates, $|\dot{M}_0|>\dot{M}_c$, only non-oscillatory runaway expansion is possible, with a critical luminosity $L_{c,dir}$, which is weakly dependent on $\dot{M}_0$. The value of $\dot{M}_c$ may depend on the specifics of the model (PNS properties, initial shock radius, dissociation losses, etc.), but for reasonable parameters appears to be about $1\;M_\odot {  s}^{-1}$.

\item We also applied the quasi-stationary approximation to examine the commonly-used ratio between the advection time scale and the heating time scale in the gain region of the accretion layer (see \S \ref{sec:explosion_condition}). We find that for the gain region to have a positive effective energy, this ratio must be at least a few tenths, with some dependence on the specific loss of energy due to dissociation across the shock (Figure \ref{fig:tadv_theat_qd0}). This result implies that the ratio of time scales, while not a fundamental condition for an explosion, can serve as an indication if a given system will evolve to an explosion, as suggested in previous works. However, we note that there are several qualifications to this observation. First, a positive energy in the gain region appears to be a threshold that must be crossed on the path to runaway expansion, yet it is not an actual condition for an explosion, since it does not exactly coincide with the region of positive shock acceleration (Figure \ref{fig:phase_space_E_lined}). Second, we find that in many cases the advection time becomes longer than the heating time only after the transition to a runaway expansion has occurred, implying that an equality between the two time scales can be a consequence of a successful explosion, rather than a physical condition for initiating one. Third, we do not identify the gain region as being unique in some physical sense; in principle, a alternative criteria can be formulated by applying it to a different region in the accretion layer, such as that which holds the largest total energy. We therefore suggest that the properties of the gain region should not be used as a singular measure of the likelihood of an explosion. 

\end{enumerate}

Since our simulations are inherently limited to spherical symmetry (with additional assumptions and simplifications), quantitative differences are to be expected in the actual problem of the neutrino driven mechanism. Nonetheless, we do believe that the qualitative nature of our results is applicable to the full, multi-dimensional problem. In particular, we emphasize the dependence of the shock acceleration on its radius and velocity; this fact certainly facilitates dynamical spherical simulations to explode with a lower neutrino luminosity than predicted by the stationary approximation. We speculate that this behavior does carry over to the three dimensional reality, where the shock can sample a wider range of combinations of $\RS$ and $\VS$ than in the one dimensional case. This should result from initial aspherical conditions in the silicon-oxygen layer \citep{couch_ott2013}, and also from the onset of turbulence (of course, turbulent pressure should also contribute to facilitating an explosion, (see \citet{couch_ott2015,Fernandez2015})). This combination may be responsible - at least in part - for the observation that the critical luminosity in two and three dimensional simulations is lower than in the one dimensional case.

A natural expansion of this work would be to examine two- and three dimensional simulations, and to quantitatively compare the actual acceleration of the shock with that predicted from a spherical model of identical parameters. Such a comparison will clarify the dependence of the shock characteristics upon the physical processes inside the accretion layer, such as turbulence and convection.

\acknowledgments

We thank E. Livne,  L. Metzker and O. Pejcha for helpful discussions. 
UK is supported by the European Union Seventh Framework Programme, by an IAEC-UPBC joint research foundation grant, and by an ISF-UGC grant.


\appendix

\newcommand{\MyTempA}{(d^2I/dt^2)_{QS}}

\section{The Derivation of $\protect\MyTempA$}\label{app:d2Idt2}

In this appendix we derive the various terms which appear in the quasi-stationary approximation for $d^2I/dt^2$ (equation \ref{eq:d2I/dt2_QS}). Recall that the goal of the approximation is to include the effect of a finite shock velocity on the profile of the accretion layer.
To do so, we assume that the profile adjusts instantaneously to changes in the shock radius and velocity. 
This requires some care, as we show below, because  the advection time is not negligible.

\subsection{Virial theorem for spherical accretion}

In general form, the first and second time derivatives of the moment of inertia in the accretion layer are:
\begin{equation} \label{eq:dI/dt}
\frac{\mathrm dI}{\mathrm d t} =
\int_{R_p}^{R_s} 8 \pi r^3 \rho u \mathrm{dr} - \left[r^2 \dot{M}(r) - 4 \pi r^4 \rho V \right]^S_P  \coma
\end{equation}
where we used the continuity equation (\ref{eq:con_density}), 
and
\begin{equation}\label{eq:d2Idt2_full}
\frac{\mathrm d^2 I}{\mathrm d^2 t}=
\int_{R_P}^{R_S} 8 \pi r^3 \frac{\mathrm \partial(\rho u)}{\mathrm \partial t} \mathrm d r -\left[r^2 \ddot{M}  -\frac{\mathrm d}{\mathrm d t} \left( 4 \pi r^4 \rho V \right)  \right]^S_P
\end{equation}
where $\dot{M}(r)=4\pi r^2 \rho u$ is the local mass accretion rate, $\ddot{M}=d \dot{M} /dt$ is its time derivative, and $V$ is the velocity of the flow boundaries. As was defined in \S \ref{sec:shock_acceleration},
the notation $[\cdots]^S_P$ stands for the difference between the expression in the immediate downstream of the shock and just outside the PNS. The last term in Equation (\ref{eq:d2Idt2_full}) is a Lagrangian time derivative of the boundaries.

Using the momentum conservation equation (\ref{eq:con_momentum}) we express $d^2I/dt^2$ in a "virial theorem" fashion with boundary conditions:
\begin{equation} \label{eq:d2I/dt2_vir}
\frac{1}{2} \frac{\mathrm d^2 I}{\mathrm d t^2}  = \mathbb{E}+\Delta\mathbb{E}_{PNS}+\Delta\mathbb{E}_S
\end{equation}
where
\begin{equation} \label{eq:energy_contribution}
\mathbb{E}=2K+3\int_{R_p}^{R_s} \frac{p}{\rho} \mathrm d m + \Omega \coma
\end{equation}
\begin{equation} \label{eq:boundary_contribution_PNS}
\Delta\mathbb{E}_{PNS}  = \left[4 \pi r^3 \left( p + \rho u^2 \right) + \frac{1}{2} r^2 \ddot{M} -\frac{\mathrm d}{\mathrm d t} \left( 2 \pi r^4 \rho V \right)\right ]_{r\rightarrow R_{P}} \coma
\end{equation}
\begin{equation} \label{eq:boundary_contribution_S}
\Delta\mathbb{E}_S=\left[-4 \pi r^3 \left( p + \rho u^2 \right) -\frac{1}{2} r^2 \ddot{M} +\frac{\mathrm d}{\mathrm d t} \left( 2 \pi r^4 \rho V \right)\right ]_{r\rightarrow R_{S}} \coma 
\end{equation}
and $K$ and $\Omega$ are the total kinetic and gravitational energies, defined as
\begin{equation}
K \equiv \int_{R_p}^{R_s} \frac{1}{2} u^2  
\mathrm{dm} 
\end{equation} 
and 
\begin{equation} \label{eq:OmegaDef}
\Omega \equiv -\int_{R_p}^{R_s} \frac{G M(m)}{r} \mathrm{dm} \approx -\int_{R_p}^{R_s} \frac{G M_p}{r} \mathrm{dm} \fin
\end{equation} 
The physical interpretation of $\mathbb{E}$ and $\Delta\mathbb{E}_{PNS}+\Delta\mathbb{E}_S$ is the energetic contributions of the accretion layer and of the boundaries (at the PNS and at the shock, respectively). We note that the boundary terms are calculated within the accretion layer, just inside the shock at $\RS$ and just above the PNS at $\RP$. The time derivative $d/dt$ in Equations (\ref{eq:boundary_contribution_PNS}-\ref{eq:boundary_contribution_S}) is a Lagrangian derivative of values in the post-shocked region, which is identically zero for a stationary PNS surface.

The $dI^2/dt^2$ expression in Eqs.~(\ref{eq:d2I/dt2_vir}--\ref{eq:OmegaDef}) is precise, but does not give accurate results in the context of a quasi-steady model.
This occurs because the $P/\rho$ term in $\mathbb{E}$ is everywhere distorted by the motion of the shock, which is nonphysical when the advection time is not negligible with respect to the characteristic time for changes in the shock velocity, \eg the oscillation time. 
Therefore, we next seek an alternative formulation that is more susceptible for the quasi-steady approximation. 

\subsection{The Effective Energy Term}

We now turn to estimate the effective energy term $\mathbb{E}$ in Equation (\ref{eq:d2I/dt2_QS}), using the quasi-stationary approximation by decomposing it into several terms. Integrating the energy-conservation equation (\ref{eq:con_energy}) over volume in a stationary profile  ($\partial/\partial t = 0$) profile yields:
\begin{equation}
\dot{M} \left[ b_S -b(r) \right] = Q(r),
\end{equation}
where $b(r)$ is the Bernoulli function (specific energy) at $r$,
\begin{equation}
b(r) \equiv \frac{1}{2} u^2 + e + \frac{p}{\rho} - \frac{G M_p}{r} \coma
\end{equation}
$b_S\equiv b(R_S)$, 
and $Q(r)$, also defined in the main text (Equation (\ref{eq:Q_r})), is the total non-adiabatic energy contribution rate between $r$ to $R_S$,
\begin{equation} \label{eq:Q_r_app}
Q(r) \equiv \int_r^{R_s} 4 \pi  \rho r^2 \dot{q} \mathrm{dr}=\int_r^{R_s} \dot{q} \, \mathrm{dm} \fin
\end{equation}
Using equation (\ref{eq:RH_energy}) for energy conservation across the shock yields
\begin{equation}
b_S = b_0 - q_d +V_S (u_1-u_0)   
= -q_d -\left( 1- \frac{1}{\beta} \right) \left( 1- \frac{V_S}{u_0} \right) u_0 V_S \coma
\end{equation}
where the Bernoulli function just outside the shock,  $b_0=0$, vanishes in the pre-shocked material (Equation \ref{eq:free-fall}). The Bernoulli function inside the post-shocked flow is then obtained by
\begin{equation} \label{eq:bernoulli}
b(r) = \frac{1}{2}u^2+e+\frac{p}{\rho} - \frac{G M_P}{r} =\frac{1}{|\dot{M} | }Q(r) +b_S \coma
\end{equation}
and the ratio $P/\rho$ in the shock can be rewritten as
\begin{equation} \label{eq:p_div_rho}
\frac{p}{\rho}=\frac{\gamma-1}{\gamma} \left( \frac{1}{|\dot{M}|}Q(r) + b_s + \frac{G M_p}{r}  -\frac{1}{2}u^2 \right) \coma
\end{equation}
where $\gamma$ is the local adiabatic index of the flow.
The total non-adiabatic energy gained by the material in the accretion layer is then
\begin{equation} \label{eq:energy_gain_U}
U = \int_{R_P}^{R_S} \frac{1}{|\dot{M}|}Q(r)  \mathrm{dm}\;.
\end{equation}

Substituting these derived quantities into the integrals included in the effective energy $\mathbb{E}$ from Equation (\ref{eq:d2I/dt2_QS}), we arrive at a form for the effective total energy in the quasi-stationary approximation, $\mathbb{E}_{QS}$
\begin{equation} \label{eq:energy_QS_long}
\mathbb{E}_{QS} = \tilde{K} + \tilde{U} + \tilde{\Omega} + \tilde{B}_S
\end{equation}
The terms in Equation (\ref{eq:energy_QS_long}) are effective manipulations of the original functions, including coefficients which depend on the local adiabatic index in the accretion layer. We define $\tilde{K}$, $\tilde{U}$, $\tilde{\Omega}$ and $\tilde{B}_S$ as the effective kinetic energy, effective gained energy, effective gravitational potential and effective Bernoulli function, respectively:
\begin{equation} \label{eq:effective_K_app}
\tilde{K} \equiv  \int_{R_p}^{R_s} \left( \frac{3-\gamma}{\gamma} \right) \frac{1}{2} u^2 \,\mathrm{dm} \coma
\end{equation}
\begin{equation} \label{eq:effective_gravity_app}
\tilde{\Omega} \equiv \int_{R_p}^{R_s} \left(\frac{3-2\gamma}{\gamma} \right)  \left( -\frac{G M_p}{r} \right) \,\mathrm{dm} \coma
\end{equation}
\begin{equation} \label{eq:effective_G_app}
\tilde{U} \equiv \int_{R_p}^{R_s} \left( 3 \frac{\gamma-1}{\gamma}  \right)  \left(\frac{1}{|\dot{M}|}Q(r) \right) \mathrm{dm}
 \coma
\end{equation}
and
\begin{equation}\label{eq:effective_B_app}
\tilde{B}_S \equiv \int_{R_p}^{R_s} \left( 3 \frac{\gamma-1}{\gamma}  \right)  b_S(m) \, \mathrm{dm}\fin
\end{equation}
The last expression requires some discussion. The quantity $b_S$ in the integral should reflect the Bernoulli function $b_S(m)$ of the mass element $m$ {\it when it crossed the shock}, and so is not a constant across the accretion layer. For example, when dissociation is neglected, $b_S = V_s(u_s-u_0)$, which vanishes in the stationary case, and fluctuates between positive and negative values as the shock oscillates. In general, this causes the integral $\tilde{B}_S$ to be largely negligible with respect to the other terms of the effective energy, which we confirm explicitly in the simulations. Hence, in the quasi-stationary approximation:
\begin{equation} \label{eq:energy_QS}
\mathbb{E}_{QS} \approx \tilde{K} + \tilde{U} + \tilde{\Omega} \;.
\end{equation}

\newcommand{\MyTempB}{(\ref{eq:acceleration_QS})}

\subsection{The Boundary Terms}

The boundary terms in $d^2I/dt^2$ includes distinct contribution from the PNS and from the shock. 
A key feature of our model is that the boundary contributions $\Delta\mathbb{E}_{PNS}$ and $\Delta\mathbb{E}_{S}$ (Eqs.~(\ref{eq:boundary_contribution_PNS}-\ref{eq:boundary_contribution_S})) are invariant under the Rankine-Hugoniot relations Eqs.~(\ref{eq:RH_density}-\ref{eq:RH_energy}). Therefore, we can determine the boundary terms using the known conditions  outside the postshocked region between the PNS and the shock.

At the PNS radius the inner boundary can be estimated with the quantities just below $R_P$, which include the predetermined pressure, $P_P$. The last term in equation (\ref{eq:boundary_contribution_PNS}) is null since $R_P$ is fixed $(V_P=0)$, and so
\begin{equation}\label{eq:W_PNS_app}
\Delta\mathbb{E}_{PNS}=4 \pi R_P^3 \left( P_P+\rho_P u_P^2 \right) + \frac{1}{2} R_P^2 \ddot{M}_P\;.
\end{equation}
The result in Equation (\ref{eq:W_PNS_app}) is general, but typically thermal pressure dominates over ram pressure at the PNS and we can approximate $P_P+\rho_P u_P^2\approx P_P$. In fact, in constructing the phase space, we found that accuracy can be increased further by imposing the fixed total pressure at the $P_P$ as the sum of the actual thermal pressure at the PNS and the initial value of $\rho_p u_P^2$, which is a correction of a few percents. Thus the approximation is only that $\rho_p u_P^2$ is not updated as the flow evolves, and this error is limited to less than one percent. In the quasi-stationary approximation $\dot{M}$ is uniform in space (although allowed to vary in time). Hence, we set $\dot{M}_P \simeq \dot{M}$ and $\ddot{M}_P  \simeq \ddot{M}$ and use equation (\ref{eq:dotM_S}).

At the shock, the quantity $\Delta\mathbb{E}_S$ is also invariant under the Rankine-Hugoniot relations can be calculated from the upstream values (index 0) instead of the shocked values (index $1$):
\begin{equation}\label{eq:W_S_shock_1st}
\begin{split}
\Delta\mathbb{E}_S  & =4 \pi R_S^3 \left( P_1+\rho_s u_1^2 \right) + \frac{1}{2} R_S^2 \ddot{M}(R_S)  - \frac{\mathrm d}{\mathrm d t} \left( 2 \pi R_S^4 \rho_s V_S \right) \\
& = 4 \pi R_S^3 \left( P_0 + \rho_0 u_0^2 \right) + \frac{1}{2} R_S^2 \ddot{M}_0  - \frac{\mathrm d}{\mathrm d t} \left( 2 \pi R_S^4 \rho_0 V_S \right)
\end{split}
\end{equation}
Finally, for accretion through the shock arriving as pressure-less free-fall at a constant accretion rate, $\dot{M}_0$ (so $\ddot{M}_0=0$), and $W_S$ is equal to:
\begin{equation} \label{eq:W_S_shock_fin}
\Delta\mathbb{E}_S=\alpha (G M_P)^{1/2}|\dot{M}_0 | \left( R_s^{1/2} - \frac{1}{7 \alpha^2 G M_P} \frac{d^2}{dt^2} \left( R_s^{7/2} \right) \right)
\end{equation}

For further application in the quasi-stationary model we denote the work done by the PNS and the energy advected across the boundaries by $W_{PNS}$ and $W_B$, respectively, as
\begin{equation} \label{eq:boundary_condition_PNS}
 W_{PNS}\simeq 4 \pi R_P^3 P_P \coma
\end{equation}
\begin{equation} \label{eq:boundary_condition_advection}
 W_B\simeq - \alpha (G M_P)^{1/2}|\dot{M}_0 |  R_S^{1/2} + \frac{ |\dot{M}_0 |}{7 \alpha (G M_P)^{1/2}} \frac{d}{dt} \left[ (\delta +1) \frac{d}{dt} \left( R_S^{7/2} \right) \right] .
\end{equation} 
The equation is approximate due to assuming a uniform accretion rate and neglecting the ram pressure. Note that under the above assumptions we maintain the sum $W_{PNS}+W_B \simeq \Delta\mathbb{E}_{PNS}+\Delta\mathbb{E}_S$.
Note that $\delta\equiv \left( \beta -1 \right) (R_P/R_S)^2$ being at the order of unity. 

It is noteworthy that the shock related terms do include an explicit dependence on the shock radius. The first term in Equation (\ref{eq:boundary_condition_advection}) is negative and restrains the shock, corresponding to the impulse of the accreting matter on the shock, while the term in Equation (\ref{eq:boundary_condition_PNS}) is positive (assisting shock expansion) and accounts for the kinetic energy of the shocked matter and the work it does on the shock.

Summarizing the results we conclude that the second time derivative of the spherical moment of inertia in the quasi-stationary approximation presented in the main text is given by:
\begin{equation} \label{eq:d2I/dt2_QS_app}
\frac{1}{2} \frac{\mathrm d^2 I}{\mathrm d t^2}  \simeq \tilde{K} + \tilde{U} +  \tilde{\Omega} +  4 \pi R_P^3 (P_P+\rho_P u_P^2) - \alpha (G M_P)^{1/2}|\dot{M}_0 |  R_S^{1/2}  + \frac{ |\dot{M}_0 |}{7 \alpha (G M_P)^{1/2}} \frac{d}{dt} \left[ (\delta +1) \frac{d}{dt} \left( R_S^{7/2} \right) \right] \;.
\end{equation}

\section{On the Importance of the Different Terms in Equation \MyTempB}\label{app:P_P}

Our final estimate for the shock acceleration in Equation (\ref{eq:acceleration_QS}) includes the effective energy of the accretion layer, $\mathbb{E}_{QS}$, the work done by the PNS, $W_{PNS}$, and two terms which depend on the properties of the shock. Here we show that $W_{PNS}$, which is invariably positive, generally dominates over the shock terms. Correspondingly, a positive $\mathbb{E}_{QS}$ can serve as a sufficient (albeit not necessary) condition for a positive shock acceleration.

First, recall that $P_P$ is calculated through a stable, stationary accretion flow which corresponds to a point on the $V_S=0$ axis and is bound such that $\mathbb{E}_{QS} < 0$. Requiring $d^2I/dt^2=0$ along with $\mathbb{E}_{QS} < 0$ in equation (\ref{eq:d2I/dt2_QS}) implies that in the stationary solution, the pressure term at the PNS must dominate over the mass influx at the shock:
\begin{equation}\label{eq:term1}
4\pi R^3_P P_P  > \alpha (G M_P)^{1/2}|\dot{M}_0|R^{1/2}_S = |\dot{M}_0 R_S u_0 |
\end{equation}
This inequality holds as long as the profile is still in the oscillatory regime. It becomes invalid once the flow has reached runaway expansion (large shock radii), but by then the explosion is assumed to be well under way with a large shock velocity, and our quasi-stationary approximation breaks down in any case.

The second term originating from the shock in Equation (\ref{eq:acceleration_QS}) is $(-V^2_S\partial\zeta/\partial R_S)$, with $\zeta$ defined in Equation (\ref{eq:zeta}). Taking the partial derivative with respect to the shock radius yields

\begin{equation} \label{eq:term2}
V^2_S\frac{\partial\zeta}{\partial R_S} \simeq V^2_S\frac{\left( \beta-1 \right)|\dot{M}_0|}
{2 \alpha (G M_P)^{1/2}}\left[\left(R_S^2-R_P^2 \right)\frac{R^{-1/2}_S}{2}+2R_S R^{-1/2}_S\right ]
= \frac{V^2_S}{u^2_0}\frac{\left(\beta-1 \right)}{2}\left(\frac{5}{2}-\frac{R^2_P}{2R^2_S}\right)|\dot{M}_0 R_S u_0|
\end{equation}

The equality is approximate because we nefglected the weak dependence of $\beta$ on the shock radius.
Now, for the small shock velocities we consider in the oscillatory phase, $V_S/u_0\lesssim <0.1$, the coefficient of $|\dot{M}_0 R_S u_0|$ in the expression in Equation (\ref{eq:term2}) is much smaller than unity ($\beta\approx 6-10$ and $\left(5/2-R^2_P/2R^2_S\right)<5/2$). Correspondingly, $V^2_S\partial\zeta/\partial R_S$ is subdominant to the first shock related term (Equation (\ref{eq:term1})), and does not change the hierarchy in which the PNS term dominates.

\section{Shock Oscillations Timescales}\label{app:t_osc}

In this appendix we show that the virial theorem can be used to explain the typical oscillation period of tens of milliseconds we find in the simulations. Coming back to Equations (\ref{eq:dIdt_simple}), (\ref{eq:d2I/dt2_vir}) and (\ref{eq:W_PNS_app}--\ref{eq:boundary_condition_advection}) we can write a oscillator-like equation for the shock radius when assuming a uniform accretion rate in the postshocked region,
\begin{equation} \label{eq:shock_energy}
\begin{split}
\frac{\mathrm d}{\mathrm d t} \left( 2 \pi \phi \rho_0  R_S^4 V_s \right) \simeq \mathbb{E} + 4 \pi R_P^3 P_P - |\dot{M}_0 R_S u_0|  \\
\end{split}
\end{equation}
where $\phi\equiv(\beta-1)[1-(R_P/R_S)^2]$ and $\mathbb{E}$ is defined in equation \ref{eq:energy_contribution} (i.e., the quasi-stationary approximation for the energy is not necessary in this context). Equation (\ref{eq:shock_energy}) was found to be in good agreement with the shock motion for both zero and full dissociation parameters. For small oscillations around a stationary solution for which $V_S=0$ at $R_{S,0}$, the right hand side must be zero at $R_{S,0}$. 

We now assume that during the oscillations, the change in the effective energy is roughly proportional to the change of the inertia crossing the shock: $\Delta\mathbb{E}=\mu |\dot{M}_0 R_S u_0|$, with $0 <\mu < 1$. This is a lowest-order approximation, in which the change in $\mathbb{E}$ depends only on the shock radius (the dependence on the shock velocity is neglected). We confirmed this assumption quantitatively in the simulations. It conveys the fact that in small oscillations the accretion layer adjusts to include the material that is either added or lost as the shock moves. Correspondingly, (recalling that $W_{PNS}$ is constant in the quasi-stationary approximation):
\begin{equation} \label{eq:shock_energy_approx}
\frac{\mathrm d}{\mathrm d t}  \left[ \frac{\phi }{2 \alpha (G M_P)^{1/2}} |\dot{M}_0|   R_S^{5/2} V_S \right]=-(1-\mu) \alpha (G M_P)^{1/2} |\dot{M}_0| \left( R_S^{1/2} - R_{S,0}^{1/2}  \right)\;.
\end{equation}
Considering that the shock radius is significantly larger than the PNS radius, $1-(R_P/R_S)^2\approx 1$ and combined with the fact that the compression ratio depends weakly on the shock radius, we can treat $\phi$ as roughly constant during the oscillations. Equation (\ref{eq:shock_energy_approx}) can now be rewritten as
\begin{equation}
\frac{\mathrm d^2 R^{7/2}_S}{\mathrm d t^2} = -\frac{ 7 (1-\mu) \alpha^2 G M_P}{\phi} \left(R^{1/2}_S-R^{1/2}_{S,0}\right)\;.
\end{equation}
It is noteworthy that the mass accretion rate $\dot{M}_0$ has canceled out of the equation. For small oscillations, this is an harmonic oscillator equation with a time period of:
\begin{equation}
T_O=2 \pi \frac{R_{S,0}}{|u_0|}  \sqrt{\frac{\phi}{1-\mu} }
\end{equation}
For a compression ratio in the range $\beta=6-10$ and $R_P/R_S<0.5$, the value of $\sqrt{\phi}$ is confined to values of $2--3$. Unless $\mu$ is very close to unity, we conclude that the oscillation period should be $(10-20)\;R_{S,0}/|u_0|$, or $(25--50)$ milliseconds. This is indeed in good agreement with the results of the simulations. We also recover the general trend that the period should be weakly dependent on the mass accretion rate and on the neutrino luminosity, as their effect is limited to the finer details of $\sqrt{\phi/(1-\mu)}$.

Finally, we note that in the approximation above, the oscillations are unconditionally stable. This is due to the neglect of a $\partial (\Delta\mathbb{E})/\partial V_S$ term in the derivation. This partial derivative is directly related to the $\partial a_S /\partial V_S$ derivative discussed in the main text, which determines whether the shock oscillations around the stationary solutions will damp or grow to a runaway expansion.\\


\clearpage

\begin{thebibliography}{}
\bibitem[Bethe(1990)]{bethe90} Bethe, H.~A. 1990, Rev.~Mod.~Phys., 62, 801
\bibitem[Bethe \& Wilson(1985)]{bethe_wilson1985} Bethe, H. A., Wilson, J. R. 1985, \apj, 295, 14
\bibitem[Bruenn et al.(2009)]{bruennetal2009} Bruenn, S. W., et al. 2009, J.~Phys.~Conf.~S., 180, 012018
\bibitem[Burrows(2013)]{Burrows2013} Burrows, A. 2013, Rev.~Mod.~Phys., 85, 345
\bibitem[Burrows \& Goshy(1993)]{burrows_goshy93} Burrows, A., Goshy, J. 1993, \apj, 416, L75
\bibitem[Burrows et al.(2007a)]{burrows2007a} Burrows, A., Dessart, L., Livne, E. 2007a, in AIP Conf. Ser. 937, Supernova 1987A: 20YearsAfter: Supernovae andGamma-Ray Bursters, ed. S. Immler, K. Weiler, \& R. McCray (Melville, NY: AIP), 370
\bibitem[Burrows et al.(2007b)]{burrows_livne07} Burrows, A., Livne, E., Dessart, L., Ott, C. D.,  Murphy, J. 2007b, \apj, 655
\bibitem[Couch(2013)]{Couch2013} Couch, S.~M. 2013, \apj, 775, 35
\bibitem[Couch \& O'connor(2014)]{couchocconor2014} Couch, S.~M. \& O'Connor, E.~P. 2014, \apj, 785, 123
\bibitem[Couch \& Ott(2013)]{couch_ott2013} Couch, S.~M. \& Ott, C.~D. 2015, \apj, 778, L7
\bibitem[Couch \& Ott(2014)]{couch_ott2014} Couch, S.~M. \& Ott, C.~D. 2014, \apj, 785, 123
\bibitem[Couch \& Ott(2015)]{couch_ott2015} Couch, S.~M. \& Ott, C.~D. 2015, \apj, 799, 5
\bibitem[Couch et al.(2015)]{couchetal2015} Couch, S.~M. Chatzopoulos, E., Arnett, D.~W. \& Timmes, F. X. 2015, arXiv:1503.02199
\bibitem[Dolence et al.(2013)]{dolenceetal2013} Dolence, J.~C., Burrows, A., Murphy, J.~W. \& Nordhaus, J. 2013, \apj, 765, 110
\bibitem[Dolence, Burrows \& Zhang(2015)]{dolenceetal_15} Dolence, J.~C., Burrows, A. \& Zhang, W. 2015, \apj, 800, 10
\bibitem[Fern\'andez(2012)]{Fernandez2012} Fern\' andez R., 2012, \apj, 749, 142
\bibitem[Fern\'andez(2015)]{Fernandez2015} Fern\' andez R., 2015, arXiv:1504.07996
\bibitem[Fern\'andez & Thompson(2009a)]{ferthom09a_dis} Fern\' andez R. \& Thompson, C., 2009, \apj, 697, 1827
\bibitem[Fern\'andez & Thompson(2009b)]{ferthom09b_rec} Fern\' andez R. \& Thompson, C., 2009, \apj, 703, 1464
\bibitem[Foglizzo et al.(2015)]{Foglizzoetal2015} Fogllizo, T. et al. 2015, \pasa, 32,9
\bibitem[Hanke et al.(2012)]{Hankeetal2012} Hanke, F., Marek,A., Muller, B., \& Janka H.-T. 2012. \apj 755, 138
\bibitem[Janka(2001)]{jan01} Janka, H.T. 2001, \aap, 368, 527
\bibitem[Janka(2012)]{Janka2012} Janka H.T., 2012, Ann. Rev. Nucl. Part. Sci., 62, 407
\bibitem[Keshet \& Balberg(2012)]{keshet_balberg12} Keshet, U., Balberg, S. 2012, \prl, 108, 251101
\bibitem[Kitaura et al.(2006)]{Kitauraetal2006} Kitaura F.~S., Janka H.-T. \& Hillebrandt W. 2006, \aap, 450, 345
\bibitem[Kushnir(2015)]{Kushnir2015} Kushnir, D. 2015, arXiv:1502.03111
\bibitem[Lentz et al.(2015)]{Lentzetal2015} Lentz, E. J. et al. 2015, arXiv:1505.05110
\bibitem[Liebendorfer et al.(2001)]{Liebendorferal2001} LiebendÃ¶rfer, M., Mezzacappa, A., Thielemann, F.-K., Messer, O.~E.; Hix, W.~R. \& Bruenn, S.~W. 2001, \prd, 63, 103004
\bibitem[Marek \& Janka(2009)]{marek_janka09} Marek, A., Janka, H.-T. 2009, \apj, 694, 664
\bibitem[Melson, Janka \& Marek(2015)]{Melsonetal2015}, Melson, T., Janka, H.-T. \& Marek, A. 2015, \apj 801L, 24
\bibitem[Murphy \& Burrows(2008)]{murphy_burrows08} Murphy, J. W., Burrows, A. 2008, \apj, 688, 1159
\bibitem[Murphy, Dolence \& Burrows(2013)]{murphyetal_13} Murphy, J.~W., Dolence, J.~C. \& Burrows, A. 2013, \apj, 771, 52
\bibitem[Murphy \& Dolence(2015)]{Murphy_Dolence2015} Murphy, J.~W. \& Dolence, J.~C. 2015 arXiv:1507.08314
\bibitem[Nagakura et al.(2013)]{nagakura13} H. Nagakura, Y. Yamamoto, S. Yamada 2013, \apj, 765, 123
\bibitem[Nordhaus et al.(2010)]{nordhaus10} Nordhaus, J., Burrows, A., Almgren, A., Bell, J. 2010, \apj, 720, 694
\bibitem[Onishi, Kotake \& Yamada(2006)]{onishietal06} Onishi, N., Kotake, K. \& Yamada, S. 2006, \apj, 641, 1018
\bibitem[Ott et al.(2013)]{Ottal2013} Ott, C.~D. et al. 2013, \apj, 768, 115
\bibitem[Pejcha \& Thompson(2012)]{pejcha_thompson12} Pejcha, O., Thompson, T. A. 2012, \apj, 746, 106
\bibitem[Richtmyer \& Morton (1967)]{richtmyer} Richtmyer, R. D., \& Morton, K. W. 1967, Difference Methods for Initial-Value Problems (2nd. ed.; New York: Interscience)
\bibitem[Sedov(1982)]{sedov} Sedov, L. I. 1982, Similarity and Dimensional Methods in Mechanics (9th ed.; Moscow: Mir Publishers)
\bibitem[Smartt(2009)]{Smartt2009} Smartt, S.~J. 2009, \araa, 47, 63S
\bibitem[Takiwaki, Kotake \& Suwa(2014)]{Takiwakietal2014} Takiwaki, T., Kotake, K.\& Suwa, Y. 2014, \apj, 786, 83
\bibitem[Wilson(1985)]{wilson1985} Wilson, J. R. 1985, in Numerical Astrophysics, ed. J. M. Centrella, J. M. LeBlanc, R. L. Bowers (Jones and Bartlett Publ., Boston), 422
\bibitem[Wongwathanarat et al.(2010)]{wong10} Wongwathanarat, A., Janka, H.-T., MÂ¨uller, E. 2010, \apj , 725, L106
\bibitem[von Neumann \& Richtmyer(1950)]{vonNeumann} von Neumann, J., \& Richtmyer, R. D 1950, J. Appl. Phys., 21, 232\bibitem[Yamasaki \& Yamada(2005)]{yamasaki_yamada05} Yamasaki, T., Yamada, S. 2005, \apj, 623, 1000
\bibitem[Yamasaki \& Yamada(2006)]{yamasaki_yamada06} Yamasaki, T., Yamada, S. 2006, \apj, 650, 291
\bibitem[Yamasaki \& Yamada(2007)]{yamasaki_yamada07} Yamasaki, T., Yamada, S. 2007, \apj, 656, 1019

\end{thebibliography}
\end{document}